\documentclass[aps,prb,twocolumn]{revtex4}
\usepackage{epsf}
\usepackage[intlimits]{amsmath}
\usepackage{amssymb}
\usepackage{graphicx}

\newcommand{\tcc}    {TlCuCl$_3$}
\newcommand{\bacu}   {BaCuSi$_2$O$_6$}

\newcommand{\Hpar}   {{\cal H}_{\parallel}}
\newcommand{\Hcoll}  {{\cal H}_{\rm coll}}
\newcommand{\Hz}     {{\cal H}_{z}}

\newcommand{\Jcoll}  {J_{\rm coll}}
\newcommand{\Jz}     {J_{z}}
\newcommand{\Jtz}    {J_{2{z}}}
\newcommand{\Jfz}    {J_{4{z}}}
\newcommand{\Jzz}    {J_{zz}}

\newcommand{\TN}     {T_{\rm N}}
\newcommand{\TKT}    {T_{\rm KT}}

\begin{document}

\title{
Reduced dimensionality in layered quantum dimer magnets: \\
Frustration vs. inhomogeneous condensates
}

\author{Oliver R\"osch}
\author{Matthias Vojta}
\affiliation{Institut f\"ur Theoretische Physik,
Universit\"at K\"oln, Z\"ulpicher Str. 77, 50937 K\"oln, Germany}
\date{November 13, 2007}

\begin{abstract}
Motivated by recent experiments on \bacu, we investigate magnetic excitations
and quantum phase transitions,
driven either by pressure or magnetic field,
of layered dimer magnets with inter-layer frustration.
We consider two scenarios,
(A) a lattice with one dimer per unit cell and perfect inter-layer frustration, and
(B) an enlarged unit cell with inequivalent layers, with and without perfect
frustration.
In all situations, the critical behavior at asymptotically low temperatures
is three-dimensional, but the corresponding crossover scale may be tiny.
Magnetic ordering in case (B) can be discussed in terms of two condensates;
remarkably, perfect frustration renders the proximity effect ineffective.
Then, the ordering transition will be generically split, with
clear signatures in measurable properties.
Using a generalized bond-operator method, we calculate the low-temperature
magnetic properties in the paramagnetic and antiferromagnetic phases.
Based on the available experimental data on \bacu, we propose that
scenario (B) with inequivalent layers and imperfect frustration is
realized in this material, likely with an additional modulation
of the inter-layer coupling along the $c$ axis.
\end{abstract}

\maketitle


\section{Introduction}

Magnetic quantum phase transitions (QPT) constitute an intense area of
current condensed matter research.\cite{hertz,moriya,millis,ssbook,rop,rmp}
Similar to classical phase transitions, the universal critical properties
of continuous QPT are determined by the symmetry of the order parameter
and the number of space dimensions, $d$, but in addition the order-parameter
dynamics plays a distinctive role.
The criterion for mean-field behavior, being $d>4$ for classical magnetic transitions,
changes to $d+z>4$ where $z$ is the dynamical exponent.
Thus, realizing quantum critical behavior beyond mean-field often
requires low-dimensional systems,
with effectively one-dimensional (1d) or two-dimensional (2d) behavior.
Such reduced dimensionality may be achieved in structured materials,
consisting of chains or planes with a weak three-dimensional (3d) coupling:
then, there exists an energy scale $E_z$ where the behavior crosses
over from 3d at low energies to 1d or 2d at higher energies.

A number of recent experiments suggested a different route to
reduced dimensionality near magnetic QPT, namely through geometric frustration.
One group of experiments are those on heavy-fermion metals,\cite{HvL,stockert,ybrhsi,rmp}
like CeCu$_{6-x}$Au$_x$ and YbRh$_2$Si$_2$,
undergoing a transition towards an antiferromagnetic metallic state,
with properties inconsistent with the standard theory of 3d magnetic QPT
in metals.\cite{hertz,moriya,millis}
Some of the proposed theoretical explanations \cite{stockert,si} are based
on the assumption of the spin fluctuations being two-dimensional.
Remarkably, indications for 2d critical fluctuations have indeed
been found in neutron scattering \cite{stockert} on CeCu$_{6-x}$Au$_x$ at the
critical concentration $x_c=0.1$.

More recently, experiments on Mott-insulating quantum paramagnets
consisting of coupled dimers of spins 1/2 have been performed.
Magnetic QPT can be driven either by application of a magnetic field or of
pressure.\cite{tlcucl,cavadini,nikuni,rueggtlcucl,sasago,bacusio,sebastian05,sebastian}
The field-driven zero-temperature transition, from a paramagnet to an antiferromagnet
with XY order at a field $H=H_{c1}$,
belongs to the universality class of the dilute Bose gas, and the
corresponding finite-temperature transition has been termed as ``Bose-Einstein
condensation of magnons''.
In the material \bacu,\cite{sasago,bacusio}
the finite-temperature transition line has been found \cite{sebastian}
to follow $T_c \propto (H-H_{c1})^\psi$, with a shift exponent $\psi=1$ characteristic of
a two-dimensional quantum critical point (QCP),
in a temperature range of $30\,{\rm mK} < T < 1\,{\rm K}$.
As \bacu\ consists of layers of Cu dimers with a fully frustrated inter-layer coupling,
i.e. a body-centered tetragonal (bct) structure of dimers, the results have been
interpreted as dimensional reduction at a QCP arising from geometric
frustration.\cite{sebastian,batista}
We have recently shown\cite{ORMV} that frustration, while in general not leading to
2d behavior at asymptotically low energies, can strongly suppress the 3d crossover
scale $E_z$. More precisely, $E_z$ is reduced to $E_z \propto \Jz^4/J^3$, compared
to an unfrustrated situation with $E_z \propto \Jz$;
here $J$ and $\Jz$ are characteristic in-plane and inter-plane coupling scales.
(We note that, at very low energies, additional ingredients like magnetic anisotropies
and coupling to nuclear spins will modify the critical behavior of a real material.)

Returning to the compound \bacu,
detailed neutron scattering experiments\cite{rueggbacusio} have revealed
the existence of at least two magnetic ``triplon'' modes in zero field,
indicating the presence of an enlarged unit cell with inequivalent dimers.
In fact, the material undergoes a structural phase transition at a temperature of 100\,K,
below which the magnetic layers are no longer equivalent.\cite{samulon}
In such a situation, the application of a field will close the gap of the lowest triplon
mode at $H_{c1}$, but the condensate will be strongly inhomogeneous along the $c$ axis,
and hence effectively two-dimensional.
An inhomogeneous condensate for fields above $H_{c1}$ has indeed been found
in a recent NMR experiment\cite{nmr} on \bacu\ which shows the existence of (at least)
two inequivalent Cu sites with distinct magnetizations.

This discussion suggests {\em two distinct routes} towards critical points with
reduced dimensionality in layered quantum magnets:
(A) equivalent layers with perfect inter-layer frustration -- this is
the scenario proposed in Ref.~\onlinecite{sebastian} and investigated theoretically
in Refs.~\onlinecite{coleman,batista,ORMV}; and
(B) inequivalent layers\cite{rueggbacusio,batista} -- in this situation perfect frustration
may still be present, but is not a required ingredient.

The purpose of this paper is twofold:
First, we give a detailed discussion of situation (A), augmenting the
report given in Ref.~\onlinecite{ORMV}.
Second, we study a model with two sets of inequivalent layers,
i.e., two dimers per unit cell, aiming at
a semi-quantitative understanding of situation (B).
We shall show that qualitative differences between the cases
with and without perfect frustration arise,
allowing for a clear-cut experimental distinction.
As in Ref.~\onlinecite{ORMV},
our general arguments are based on an analysis of order-parameter field theories,
and quantitative calculations are performed using variants of the bond-operator
approach, in both the paramagnetic and antiferromagnetic phases.
We shall be primarily interested in the mechanisms which suppress the 3d crossover
scale to small values. The physics below this scale shall not be addressed in detail,
as e.g. material-dependent anisotropies (not included in our analysis) will
become important.

\begin{figure}[t]
\epsfxsize=1.1in
\epsffile{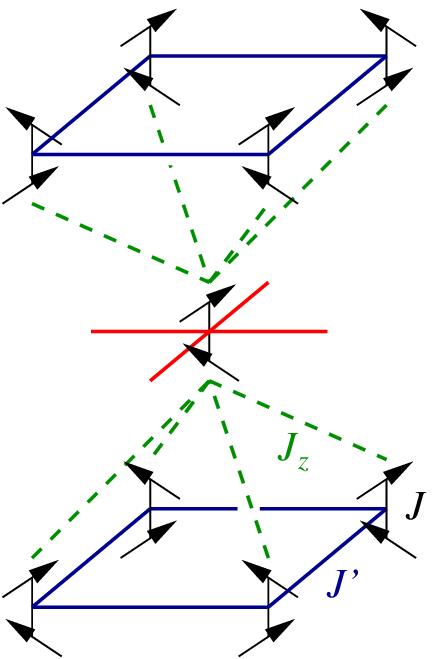}
\epsfxsize=1.1in
\epsffile{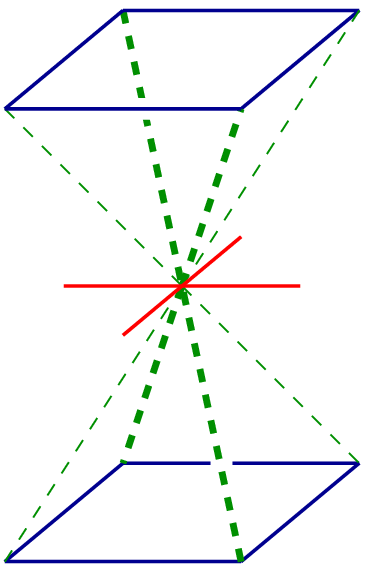}
\epsfxsize=1.1in
\epsffile{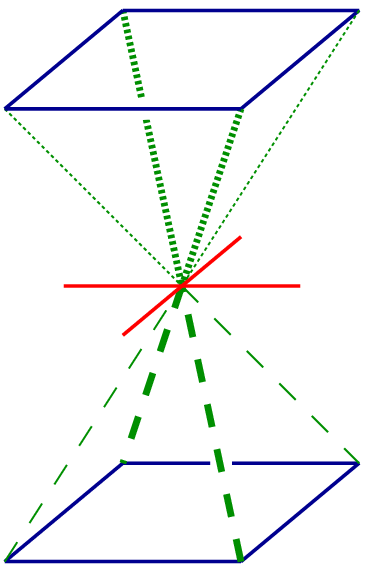}
\caption{
a) bct lattice structure of dimers, with couplings
$J$ (intra-dimer), $J'$ (intra-layer), and $\Jz$ (inter-layer).
b,c) Illustrations of possible symmetry breaking.
In b) the inter-layer couplings $\Jz^\Delta$ are modulated
within a unit cell such that the two diagonals are inequivalent.
In c) the inter-layer couplings are further modulated along
the $c$ axis, enlarging the unit cell.
}
\label{fig:struct}
\end{figure}

\subsection{Main results}

In the following, we summarize our main results which strictly apply
to a bct system of dimers (Fig.~\ref{fig:struct}), but apply with
minor modifications in general to magnetic systems with frustrated
inter-layer interaction:

(i) At the magnetic quantum critical points, the asymptotic low-energy physics
below a scale $E_z$ is in general three-dimensional.
However, the crossover scale $E_z$ is strongly reduced compared to
unfrustrated layered systems, due to either
(A) frustration, or
(B) a layer ``mismatch'' arising from inequivalent layers.
(An exception is the high-field quantum phase transition between
a canted and a fully polarized state at field $H_{c2}$,
where perfect frustration renders the low-energy physics
two-dimensional, due to the lack of scattering processes at $T=0$.\cite{batista})

(ii) Perfect frustration implies the absence of a linear coupling between two
distinct magnetic condensates (on the even and odd layers),
leading to an additional Z$_2$ symmetry which is spontaneoulsy broken
in the ordered phases.
For inequivalent layers, the two magnetic condensates will be established at
different critical points, i.e., the magnetic transition is generically split due
to the absence of a proximity effect.
This feature persists to finite temperatures and allows a distinction
between perfect and imperfect frustration --
in the latter case, the secondary transition is smeared out.

\begin{figure}[t]
\epsfxsize=3in
\centerline{\epsffile{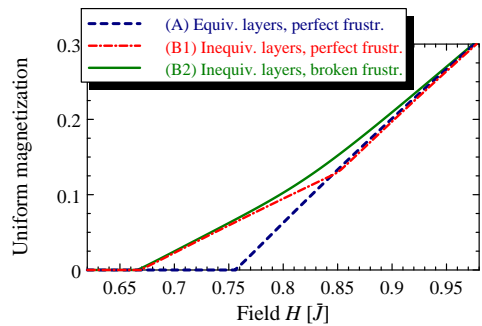}}
\caption{
Zero-temperature magnetization vs. applied field in the coupled-dimer
model near $H_{c1}$ at $T=0$,
calculated using bond operators (Sec.~\ref{sec:bond}),
for three scenarios:
(A) equivalent layers and perfect frustration (dashed),
(B1) inequivalent layers and perfect frustration (dash-dot),
(B2) inequivalent layers and imperfect frustration (solid).
Case (B1) clearly shows the secondary transition as a kink in the
magnetization curve, whereas this feature is smeared out due to
the proximity effect in case (B2).
Details will be discussed in Sec.~\ref{sec:bacu},
the parameter values are those in Figs.~\ref{fig:mag11},\ref{fig:mag12},\ref{fig:mag13}
below.
We caution the reader that logarithmic corrections, not captured by our approximation,
may somewhat modify the shape of the magnetization curve close to the critical field,
but the existence of a sharp kink in (B1) will be unaffected.
}
\label{fig:mag}
\end{figure}

(iii) Applied to the field-driven QPT in \bacu,
our results suggest that scenario (B), with inequivalent layers
and incomplete frustration, is closest to the experimental situation.
To demonstrate the qualitative differences,
we plot in Fig.~\ref{fig:mag} the magnetization curves close to $H_{c1}$ for
three prominent situations (equivalent layers with perfect frustration,
inequivalent layers with and without perfect frustration).
Finally, to reconcile the rather large magnetic proximity effect as evidenced by
NMR\cite{nmr} with the tiny 3d crossover scale\cite{sebastian}
we propose a vertically modulated inter-layer coupling
which leads to a further suppression of 3d behavior.

We note that a recent paper by Batista {\em et al.}\cite{batista}
studied in detail a hard-core boson model on a bct lattice --
this is the appropriate model at the $H_{c2}$ transition, and
approximately applies also in the vicinity of $H_{c1}$, however,
there it does not capture the interaction-generated 3d dispersion.

\subsection{Outline}

The bulk of the paper is organized as follows:
In Sec.~\ref{sec:model} we introduce the coupled-dimer model
to be studied in this paper, and discuss ways of breaking the
bct lattice symmetries.
In Sec.~\ref{sec:sym} we develop an order-parameter description of the magnetically
ordered phases, and present a comprehensive symmetry analysis.
For quantitative calculations at zero temperature,
we employ the bond-operator method, summarized in Sec.~\ref{sec:bond}.
Secs.~\ref{sec:eq} and \ref{sec:ineq} contain the central results of our paper,
both for scenario (A) with equivalent layers and full frustration (Sec.~\ref{sec:eq})
and for scenario (B) with inequivalent layers (Sec.~\ref{sec:ineq}).
We discuss the structure of the phase diagram, the nature of the phase
transitions, various energy scales relevant for thermodynamics as well as for the
magnetic excitations, and zero-temperature observables like e.g. magnetizations.
Finally, Sec.~\ref{sec:bacu} applies the results to \bacu,
by deriving theoretical constraints from various experimental results:
This leads us to propose a scenario of inequivalent layers with partially frustrated
and vertically modulated inter-layer interaction.
A brief outlook concludes the paper.

Readers primarily interested in our conclusions regarding \bacu\ may directly
jump ahead to Sec.~\ref{sec:bacu}
(after having glanced at the notations in Sec.~\ref{sec:model}).


\section{Model}
\label{sec:model}

In this paper, we will concentrate on bct lattices, Fig.~\ref{fig:struct},
consisting of two interpenetrating tetragonal subsystems.
Assuming magnetic moments to be
located on the sites of this lattice, with nearest-neighbor antiferromagnetic
couplings, the in-plane order will be N\'{e}el-like, but the coupling between
adjacent layers is fully frustrated.
The geometric frustration can be seen at the single-particle level:
For a tight-binding model with nearest-neighbor couplings
$t$ (intra-layer) and $t_z$ (inter-layer), the single-particle
dispersion is given by
$\epsilon_{\vec q} = 2 t (\cos q_x + \cos q_y) + 4 t_z \cos (q_x/2) \cos (q_y/2) \cos q_z$.
For positive $t$ and small $t_z$, the minimum of the dispersion is at
in-plane wavevector ${\vec q}_\parallel = (\pi,\pi)$, where it is {\it independent} of $q_z$ --
this is the result of inter-layer frustration.

With the application to coupled-dimer materials like \bacu\ in mind,
we will consider a Heisenberg Hamiltonian for
dimers of spins $1/2$. We decompose the Hamiltonian into in-plane and inter-plane
parts,
\begin{equation}
{\cal H} = \Hpar + \Hz.
\label{H}
\end{equation}
The in-plane part is unfrustrated and reads
\begin{eqnarray}
{\cal H}_\parallel &=&
                           \sum_{in} J_n {\vec S}_{in1} \!\cdot\! {\vec S}_{in2}
+ \! \sum_{\langle ij\rangle nm} \! J'_n {\vec S}_{inm} \!\cdot\! {\vec S}_{jnm} \nonumber \\
&-& {\vec H}\!\cdot\! \sum_{inm} {\vec S}_{inm}
\label{Hpar}
\end{eqnarray}
where $m=1,2$ labels the spins of each dimer, $i,j$ are the dimer site indices in each
layer, and $n$ the layer index.
$J_n$ and $J'_n$ are the antiferromagnetic intra-dimer and in-plane inter-dimer couplings,
and $\vec H$ is an external uniform field.

A rather general form of the inter-layer coupling reads
\begin{eqnarray}
\Hz &=&
  \!\!\sum_{i\Delta n m m'} \!\!\! \Jz^{nmm'\Delta}{\vec S}_{inm} \!\cdot\! {\vec S}_{i+\Delta,n+1,m'}
  \nonumber\\
  &+&  \sum_{inmm'} \! \Jzz^{nmm'}
{\vec S}_{inm} \!\cdot\! {\vec S}_{i,n+2,m'}.
\label{Hz}
\end{eqnarray}
$\Jz$ is the frustrated coupling between adjacent layers,
whereas $\Jzz$ represents an {\em unfrustrated} coupling between second-neighbor layers.
The $\sum_\Delta$ runs over four spatial diagonals ($\Delta=1,\ldots,4$)
such that the sites $(in)$ and
$(i+\Delta,n+1)$ are nearest neighbors in $z$ direction, i.e.,
with distance $(\pm a/2,\pm a/2,c)$ where $a$ and $c$ are the lattice constants
(which are set to unity in the following).
Note that $\Jz^{nmm'\Delta}$ has some specific inter-dimer structure given by the $mm'$
dependence, and frustration can be broken due to the $\Delta$ dependence, see below.
The $mm'$ dependence requires discussion, as different physical processes
are determined by different combinations of $\Jz^{mm'}$:
The ``bare dispersion'' of the triplet excitations is determined by
$\Jtz = \Jz^{11} + \Jz^{22} - \Jz^{12} - \Jz^{21}$, whereas the
combination
$\Jfz = \Jz^{11} + \Jz^{22} + \Jz^{12} + \Jz^{21}$
enters in the interaction vertex.
If we label the lower (upper) spin in each dimer with 1 (2),
Fig.~\ref{fig:struct}a, the geometry suggests that $\Jz^{21}$ is dominant and
antiferromagnetic.
Hence, in what follows we shall assume $-\Jtz=\Jfz\equiv \Jz > 0$ unless otherwise
noted.

Occasionally we will refer to a possible biquadratic inter-layer exchange term:
\begin{equation}
{\cal H}_{\rm coll} =
  \sum_{i\Delta n m m'} \!\! \Jcoll^{nmm'\Delta}
  ({\vec S}_{inm} \!\cdot\! {\vec S}_{i+\Delta,n+1,m'})^2.
\label{Hcoll}
\end{equation}

As we have shown in Ref.~\onlinecite{ORMV}, an effective second-neighbor
inter-layer coupling is always generated ($\propto\Jz^4$) through interaction processes,
even if a ``bare'' $\Jzz$ is absent in an idealized model Hamiltonian.
(The same applies to the biquadratic term $\Jcoll$, with $\Jcoll\propto-\Jz^2$.)
In the microscopic calculations to be presented in this paper,
we set $\Jzz=\Jcoll=0$ unless otherwise noted.

\subsection{Equivalent layers}
\label{sec:modeq}

For an ideal bct lattice structure, all layers are equivalent,
$J_n = J$, $J'_n = J'$,
and the inter-layer coupling is fully frustrated, i.e.,
the four diagonal bonds are equal,
$\Jz^{nmm'\Delta} = \Jz^{mm'}$.
Importantly, the second-neighbor coupling in $z$ direction is allowed by
symmetry,\cite{ORMV} $\Jzz \neq 0$ in general.

Distortions of the ideal bct lattice may or may not enlarge the unit cell.
In the latter case, particularly interesting are distortions
which break the perfect inter-layer frustration.
Those correspond to a $\Delta$ dependence of $\Jz^{nmm'\Delta}$
which renders the four links between one site and its neighbors in $z$ direction
inequivalent.
The simplest symmetry breaking leads to different couplings along the two diagonals,
see Fig.~\ref{fig:struct}b.

\subsection{Inequivalent layers}
\label{sec:modineq}

Various lattice distortions can occur which enlarge the unit cell.
This is likely the case in the low-temperature phase of \bacu, which, however,
to our knowledge, has not been fully characterized to date.
For simplicity and motivated by the \bacu\ neutron scattering results,\cite{rueggbacusio}
we will assume that the distortions preserve the tendency towards commensurate
in-plane ordering at wavevector $(\pi,\pi$),
but we allow for layer-dependent in-plane couplings $J$, $J'$
and possibly broken inter-layer frustration.

Importantly,
{\em within} the two sets of ``even'' and ``odd'' layers (i.e. layers with even and odd $n$)
a unfrustrated (albeit small) coupling will be present through $\Jzz$,
irrespective of the presence or absence of inter-layer frustration.
Thus the simplest and physically most interesting scenario is one where
all even planes are identical, as are the odd ones, i.e.,
we have a lattice still consisting of two tetragonal subsystems,
labeled $A$ ($n$ even) and $B$ ($n$ odd).
Then the couplings take the form:
\begin{equation}
J_{2n} = J_A,~ J'_{2n} = J'_A,~ J_{2n+1} = J_B,~ J'_{2n+1} = J'_B.
\end{equation}
This also implies two different second-neighbor vertical couplings,
$\Jzz^{2n} = \Jzz^A$, $\Jzz^{2n+1} = \Jzz^B$.

The coupling between adjacent layers may still have the full symmetry,
$\Jz^{nmm'\Delta} = \Jz^{mm'}$,
implying perfect frustration.
Alternatively, the distortion can break the frustration.
The simplest situation with inequivalent diagonals,
Fig.~\ref{fig:struct}b,
can be described by $\Jz$ couplings according to
\begin{eqnarray}
\Jz^{2nmm'\Delta}    &=& \left\{
\begin{array}{cc}
  J_{zA1}^{mm'} & \Delta~{\rm odd} \\
  J_{zA2}^{mm'} & \Delta~{\rm even}
\end{array},\right.\nonumber\\
\Jz^{2n+1,mm'\Delta}  &=& \left\{
\begin{array}{cc}
  J_{zB1}^{mm'} & \Delta~{\rm odd} \\
  J_{zB2}^{mm'} & \Delta~{\rm even}
\end{array}.\right.
\label{jzdef}
\end{eqnarray}
Here we have also allowed for a modulation of the $\Jz$ coupling along the $c$ axis,
as illustrated in Fig.~\ref{fig:struct}c -- this will turn out to be
relevant for our analysis of \bacu.
(Note that $J_A\neq J_B$, $J'_A\neq J'_B$ on the one hand
and $J_{zA}\neq J_{zB}$ on the other hand break {\em different}
mirror symmetries of the bct lattice.)

All scenarios included above lead to a unit cell containing no more than two dimers
(one from the even and one from the odd layers).
A further enlarged unit cell, i.e., multiple dimers within each family of layers,
only lead to quantitative modifications of the overall physics,
and will be discussed towards the end of the paper.

\subsection{Phases}

To set the stage, we sketch the phase diagram of the coupled-dimer model $\cal H$
(ignoring the complications arising from structural distortions of the ideal bct
lattice). At zero temperature the phase diagram is similar to that
of the well-studied bilayer Heisenberg model,\cite{sandvik,SaChuSa,kotov,troyer,sommer}
see e.g. Fig. 1 of Ref.~\onlinecite{troyer}.

For $J\gg J',|\Jz|$, the zero-field ground state of $\cal H$ is a paramagnetic singlet,
with elementary gapped triplet excitations.
The number of excitation branches is equal to the number of dimers per unit cell.

If $J'$ dominates, an antiferromagnetic phase with broken SU(2) symmetry and
in-plane ordering wavevector $(\pi,\pi)$ is established.
In the case of perfect inter-layer frustration,
layers with classical moments would be uncoupled, leaving an infinite ground-state
manifold. This degeneracy is lifted for quantum spins by a zero-temperature
order-from-disorder mechanism, which establishes true 3d order (Sec.~\ref{sec:eqofd}).
In contrast, for large $|\Jz|$ the in-plane order is ferromagnetic, with
the inter-plane order depending on the sign of $\Jz$.

Applying a field to the large-$J$ quantum paramagnet leads to a Zeeman
splitting of the triplet excitations.
At a critical field $H_{c1}$, the gap of the lowest mode closes, and
a quantum phase transition to a gapless canted phase occurs.
(For modifications due to inequivalent dimers see Sec.~\ref{sec:ineq}.)
The canted phase has a broken U(1) symmetry, corresponding to
XY order perpendicular to the field.
Upon further increasing the field, the system is driven into
a fully polarized state at $H_{c2}$.


\section{Symmetries and order-parameter field theory}
\label{sec:sym}

In this section, we present the ingredients for a Landau-Ginzburg-Wilson
description of the degrees of freedom of the coupled-dimer model $\cal H$.
We start with antiferromagnetic fluctuations on a paramagnetic background.

A remark of caution is in order:
The formal derivation of an effective theory for the staggered
magnetization involves integrating out the degrees of freedom
associated with the uniform magnetization,
see e.g. Ref.~\onlinecite{ssbook}.
(The ``local'' staggered and uniform magnetizations correspond to
$\vec S_1 - \vec S_2$ and $\vec S_1 + \vec S_2$, respectively, in terms
of the two spins $\vec S_1$, $\vec S_2$ of a dimer in $\cal H$ \eqref{H}.)
However, there exist processes which are strictly forbidden
within the effective theory for the staggered magnetization,
but exist in the full theory.
One example is the two-particle (as opposed to three-particle)
decay of triplons discussed recently in Refs.~\onlinecite{decay1,decay2};
another example is the effective vertical second-neighbor hopping
of order $\Jz^4$ in the model $\cal H$ \eqref{H},
to be described below in Sec.~\ref{sec:eqpara}.

\subsection{Antiferromagnetic order parameter}

A continuum description of antiferromagnetic fluctuations on bct lattices
requires care due to the geometric frustration.
In-plane magnetic fluctuations are strong and centered around wavevector
$\vec Q=(\pi,\pi)$, but
the effective inter-layer coupling is frustrated and weak.
Hence there will be a large regime of energies or temperatures with
no well-defined order along the $c$ axis.
Therefore we only take the continuum limit w.r.t. the in-plane coordinates,
but keep the discrete layer index $n$.
We thus define an order parameter
${\vec \phi}_n(\vec r_\parallel)$,
where $\vec r_\parallel$ is the in-plane coordinate, with the local magnetization operator given by
${\vec m}_n({\vec r_\parallel}) = \exp(i {\vec Q \cdot \vec r_\parallel}) {\vec \phi}_n(\vec r_\parallel)$.
As outlined above, we will allow for unequal layers, but assume a two-dimer unit cell.
Hence, we will use the labels $A$ ($B$) for even (odd) layers and the respective parameters.

The full $\phi^4$ theory for the magnetic fluctuations is
\begin{equation}
\label{sphi}
{\cal S}_{\phi} =
{\cal S}_{2A} + {\cal S}_{2B} + {\cal S}_{4A} + {\cal S}_{4B} + {\cal S}_{\rm dyn}
+ {\cal S}_{AB}.
\end{equation}
The first two terms, ${\cal S}_{2A} + {\cal S}_{2B}$, contain the Gaussian
description of the unfrustrated magnetism of the two tetragonal subsystems:
\begin{eqnarray}
{\cal S}_{2A} &=&
\int d\tau d^2 k_\parallel \sum_{n~\rm even} \Big[
( m_{\phi A} + c_A^2 k_\parallel^2) {\vec \phi}_{nA}^2({\vec k_\parallel})  \nonumber\\
&& ~~~~~~~~~~~~~~~~~~~~~~~~+ \eta'_A {\vec \phi}_{nA}\cdot{\vec \phi}_{n+2,A} \Big],
\label{phi4}
\end{eqnarray}
the action ${\cal S}_{2B}$ for odd layers is obtained by $A\to B$.
Here ${\vec \phi}({\vec k_\parallel})$ is the order-parameter field after
in-plane Fourier transformation, the momentum $\vec k_\parallel$ is now measured
relative to the ordering wavevector $(\pi,\pi)$, and $c$ is a velocity.
Further, $\eta'$ represents the unfrustrated vertical second-neighbor coupling.
From the microscopic model one reads off the bare value $\eta' \sim \Jzz$;
however, as shown in Ref.~\onlinecite{ORMV},
interaction effects cause a non-zero $\eta'$ even for vanishing bare $\Jzz$,
see Sec.~\ref{sec:eqpara}.
The mass $m_{\phi}$ is the control parameter of the zero-field transition;
a negative (renormalized) mass $m_{\phi A}$ leads to a magnetic condensate with
$\langle\vec\phi_A\rangle\neq 0$.
(The coupling between $\vec\phi_A$ and $\vec\phi_B$ will be discussed below.)
The terms ${\cal S}_{4A}$, ${\cal S}_{4B}$ contain the local quartic self-interaction,
\begin{equation}
{\cal S}_{4A} = u_0 \int d\tau d^2 r_\parallel \sum_{n~\rm even}
[{\vec \phi}_{nA}^2(\vec r_\parallel)]^2 \,.
\end{equation}
Finally, ${\cal S}_{\rm dyn}$ encodes the dynamics of the spin fluctuations:\cite{ssbook}
\begin{equation}
{\cal S}_{\rm dyn} =
\int d\tau d^2r_\parallel \sum_{n}
(\partial_\tau {\vec\phi}_n - i \vec H \times {\vec\phi}_n)^2 .
\label{sdyn}
\end{equation}
In zero field, there is only a second-order time derivative, the dynamical exponent is $z=1$,
and the modes are triply degenerate.
In contrast, in finite field we have $z=2$, and the modes are Zeeman-split
according to $\omega \to \omega - \alpha H$, where $\alpha=+,0,-$ and
the $+,-$ modes correspond to $\phi_x \pm i \phi_y$
(assuming the field to be in $z$ direction).

Let us now turn to the frustrated coupling between adjacent layers, described by ${\cal S}_{AB}$.
To this end, we repeat the central symmetry argument for the perfectly
frustrated geometry, already given in Ref.~\onlinecite{ORMV}.
Apart from spin rotation and space inversion symmetry,
the system is also invariant under 90-degree in-plane rotations,
however, the geometry dictates that this is accompanied by a
relative sign change of the order parameter in two neighboring planes (!):
\begin{equation}
k_x \rightarrow k_y\,,~
k_y \rightarrow -k_x\,,~
{\vec \phi}_n \rightarrow (-1)^n {\vec \phi}_n \,.
\label{sym}
\end{equation}
This symmetry strongly constrains the allowed inter-layer coupling terms
in the case of perfect frustration.
A general form of the inter-layer coupling, including quadratic and quartic terms, is
\begin{eqnarray}
{\cal S}_{AB} &=& \int d\tau d^2k_\parallel \sum_n \big[
  \kappa \vec\phi_{n} \cdot \vec\phi_{n+1} +
\eta k_x k_y \vec\phi_{n} \cdot \vec\phi_{n+1}
 \nonumber\\
&+&
u_{1} (\vec\phi_{n} \cdot \vec\phi_{n+1})^2 +
u_{2} \vec\phi_{n}^2 \vec\phi_{n+1}^2
\big].
\label{sab}
\end{eqnarray}
(To avoid clutter of notation, we have assumed the couplings to be
vertically unmodulated; a corresponding generalization is straightforward.)
Clearly, the $\kappa$ term -- actually corresponding to unfrustrated inter-layer coupling --
is {\em incompatible} with the symmetry \eqref{sym},
whereas the other terms are compatible.
Thus, perfect frustration implies $\kappa = 0$.
The $\eta$ term represents single-particle hopping in the presence of frustration,
i.e., it vanishes for $k_x=0$ or $k_y=0$;
the form of this hopping term can be directly obtained from
expanding the tight-binding dispersion on the bct lattice near in-plane wavevector $(\pi,\pi)$.
From the microscopic model \eqref{H} one reads off $\eta \propto \Jz$.

Further, $u_1$ and $u_2$ represent density interactions between adjacent layers.
An important role -- in particular in the ordered phase --
is played by the $u_1$ term:
in the presence of perfect frustration this is the leading coupling between adjacent layers
at $k_\parallel = 0$ which is sensitive to spin directions.
Negative $u_1$ stabilizes collinear spin correlations between adjacent planes,
while positive $u_1$ favors orthogonal $\vec\phi_A$ and $\vec\phi_B$.
Comparing with the microscopic model \eqref{H},
it is obvious that the $u_1$ term only arises at order $\Jz^2$,
in fact $u_1 \propto -\Jz^2/J$, see Sec.~\ref{sec:eqpara}.
(These statements hold in the absence of a bare $\Hcoll$,
otherwise $u_1 \propto \Jcoll$ dominates.)


\subsection{Full frustration: Z$_2$ symmetry and Ising bond order parameter}
\label{sec:z2op}

As the bilinear magnetic coupling between adjacent layers,
$\vec\phi_n \cdot \vec\phi_{n+1}$,
is suppressed by a prefactor of $k_x k_y$ in the presence of perfect frustration,
there is {\em no} linear coupling between the condensates
$\langle\vec\phi_A\rangle$ and $\langle\vec\phi_B\rangle$ on the even and odd layers
in an antiferromagnetically ordered phase.
Instead, the dominant $A$--$B$ coupling is given by the biquadratic term $\propto u_1$,
which will select collinear or orthogonal correlations between
the two condensates, but will always leave a Z$_2$ degeneracy intact,
corresponding to a spin inversion in every second plane.
This Z$_2$ symmetry corresponds to a true symmetry for the
antiferromagnet on the bct lattice,
and will be spontaneously broken in the ordered phase.

For an undistorted lattice,
it is then useful to introduce a local Ising order parameter $\Psi_{n+1/2}$
which is conjugate to $\vec\phi_n \cdot \vec\phi_{n+1}$ and
lives at zero in-plane wavevector.
$\Psi$ transforms as a singlet under SU(2) spin rotations, and hence
can be described by an unfrustrated $\Psi^4$ theory.
Assuming a single-dimer unit cell, a plausible form is:
\begin{eqnarray}
{\cal S}_\psi &=&
\int d\tau d^2 k_\parallel \Big[
\sum_n ( m_\Psi + c_\Psi^2 k_\parallel^2) \Psi_{n+1/2}^2({\vec k_\parallel})  \nonumber\\
&&+ \eta_\Psi \Psi_{n-1/2} \Psi_{n+1/2}
\Big]
+ {\cal S}_{\Psi 4} + {\cal S}_{\Psi \rm dyn},
\label{psi4}
\end{eqnarray}
where ${\cal S}_{\Psi 4}$ is again a quartic self-interaction, and the dynamic term
${\cal S}_{\Psi \rm dyn}$ contains a second-order time derivative.
The physical content of $\Psi$ is encoded in its interaction with $\vec\phi$,
where the leading term is trilinear:
\begin{eqnarray}
{\cal S}_{\phi\Psi} &=& \lambda
\int d\tau d^2 k_\parallel
\sum_n  \Psi_{n+1/2} \, \vec\phi_n \cdot \vec\phi_{n+1}
\label{phipsi}
\end{eqnarray}
with $\lambda$ a coupling constant.
(Additional couplings $\Psi \vec\phi_n^2$ etc.
do not modify the physics to be discussed below.)

Eq.~\eqref{phipsi} shows that the condensation of $\Psi$ induces an
{\em unfrustrated} vertical hopping through the term
$\lambda\langle\Psi\rangle\vec\phi_n \cdot \vec\phi_{n+1}$,
i.e., it breaks the frustration.
Microscopically, $\Psi$ condensation is equivalent to spontaneous bond order,
modulating the vertical magnetic couplings $\Jz^\Delta$ {\it within} each unit cell,
as illustrated in Fig.~\ref{fig:struct}b.
Within a purely magnetic model, $\Psi$ can be understood
as a singlet bound state of two $\vec\phi$ quanta,
implying that $m_\Psi$ is essentially given by $2 m_\phi$, plus
a correction arising due to an attraction or repulsion of $\vec\phi$ quanta
from ${\cal S}_{AB}$ \eqref{sab}.

In the presence of phononic degrees of freedom, bond order causes lattice
distortions, hence $\Psi$ condensation is a structural phase transition.
$\Psi$ may condense either uniformly or with a non-trivial modulation
along the $c$ axis.
(In principle, the action ${\cal S}_\Psi$ could be {\em dominated} by phonon
effects -- this will not be considered.)

Let us now discuss the implications for the phase diagram of
the spin model on the ideal bct lattice:
The full theory ${\cal S}_\phi+{\cal S}_\Psi+{\cal S}_{\phi\Psi}$
admits two distinct scenarios:
(i) A single transition driven by the condensation of $\phi$ --
here, the coupling $\lambda$ generates a non-zero expectation value for $\Psi$ as well,
because the $u_1$ term in ${\cal S}_{AB}$ leads to non-zero
$\langle \vec\phi_n\cdot\vec\phi_{n+1} \rangle$.
(ii) Two transitions: First, $\Psi$ condenses, which modifies the quadratic part
of the $\phi$ action, relieving the frustration,
and $\phi$ orders in a second, subsequent transition.
For the microscopically relevant parameters, we find that situation (i) is
generically realized, see Sec.~\ref{sec:bound}.

\subsection{Canted magnetism near $H_{c1}$}

Near the critical field $H_{c1}$, only the lowest of the Zeeman-split triplet modes
is relevant for the low-energy behavior.
This lowest mode is $\Phi = \phi_x+i\phi_y$,
and the critical theory can be formulated using a single complex scalar field $\Phi$
with canonical boson dynamics and a mass (i.e. chemical potential) $\sim (H_{c1}-H)$.
Condensation of $\Phi$ breaks a U(1) symmetry and
leads to spontaneous order perpendicular to the field direction:
the transverse staggered magnetization is given by $\langle\Phi\rangle$,
whereas the longitudinal uniform magnetization is $\langle\Phi^*\Phi\rangle$.
Thus, the physics near $H_{c1}$ is that of a dilute Bose gas.\cite{ssbook}
(Complications again arise from the additional Z$_2$ symmetry,
the detailed discussion will be given in Sec.~\ref{sec:eqhpd}.)

The above symmetry analysis in terms of $\vec\phi$ continues to apply, with the change
that $\vec\phi$ is now a two-component vector representing the transverse staggered
magnetization.
However, care is needed when associating effective with microscopic couplings,
as e.g. $\vec\phi^2$ now carries a uniform magnetization and hence couples linearly
to a field.
(Technically, this arises because the degrees of freedom of the uniform magnetization
are no longer gapped.)

\subsection{Canted magnetism near $H_{c2}$}

At high fields the ground state of the system is fully polarized,
and the elementary excitations are bosonic spin-flip quasiparticles.
Upon decreasing the field, those will condense at $H_{c2}$,
leading to a canted phase which is continuously connected to the
canted phase established above $H_{c1}$.
Hence, the order-parameter description is identical to the one near $H_{c1}$:
The order parameter is a canonical boson $\Phi$, now
with a mass $\sim (H-H_{c2})$.
The transverse staggered magnetization is again given by $\langle\Phi\rangle$,
whereas the uniform magnetization is $M_{\rm sat} - \langle\Phi^*\Phi\rangle$.

\subsection{Phase transitions}

Most of the quantum phase transitions discussed below are at or above their
upper-critical dimension
(which the exception of the Ising and O(3) transitions in $d=2$),
thus the critical exponents are known.
The shift exponent $\psi$ of the finite-temperature phase boundary
is given by the product of correlation length and dynamical exponents,
$\nu z$, if the QPT is below its upper critical dimension;
otherwise it can be obtained from the temperature dependence of
the Hartree diagram determining the mass shift of the order parameter,\cite{ssbook}
and is given by $\psi=z/(d+z-2)$ ($z=2$ for a BEC transition).

In this paper, we intend to estimate relevant energy scales
and to obtain the overall behavior of observables,
primarily at zero temperature away from the phase transitions.
To this end, we employ bare as well as self-consistent perturbation theory.
These methods may break down near criticality, and we comment on this below.

The finite-$T$ regime close to the ordering temperature is more difficult:
The interplay of the frustration-related order-from-disorder mechanism
and the Mermin-Wagner theorem is delicate, see Sec.~\ref{sec:tn}.
A reliable treatment of the finite-temperature transitions,
including an estimate of $\TN$, is beyond the scope of the paper.


\section{Bond-operator theory}
\label{sec:bond}

For a quantitative study of the coupled-dimer Heisenberg model (\ref{H})
we apply the bond-operator approach of Sachdev and Bhatt,\cite{bondop}
with extensions proposed by Kotov {\em et al.}\cite{kotov}
and by Sommer {\em et al.}\cite{sommer}
While this method can in principle be applied at finite
temperatures,\cite{bofinitet} we will restrict the explicit calculations to $T=0$.

In this section, we present the formalism for a situation with
equivalent dimers, i.e. a bct lattice with a single-site unit cell
(with lattice sites denoted by $i$).
The generalization to inequivalent dimers is straightforward,
and we shall refrain from showing the lengthy equations.

\subsection{Paramagnetic phase: Harmonic approximation}

The four states of a dimer $i$ can be represented using
bosonic ``bond'' operators $\{s_i^\dagger,t^\dagger_{i\alpha}\}$ ($\alpha=x,y,z$),
which create the dimer states out of a fictitious vacuum.
Explicitly (and omitting the site index $i$),
$|s\rangle = s^\dagger |0\rangle$,
$|\alpha\rangle = t^\dagger_{\alpha}|0\rangle$, where
$\left|s\rangle\right. =
(\left|\uparrow\downarrow\rangle\right.-\left|\downarrow\uparrow\rangle\right.)/\sqrt{2}$,
$\left|x\rangle\right. =
(-\left|\uparrow\uparrow\rangle\right.+\left|\downarrow\downarrow\rangle\right.)/\sqrt{2}$,
$\left|y\rangle\right. =
i(\left|\uparrow\uparrow\rangle\right.+\left|\downarrow\downarrow\rangle\right.)/\sqrt{2}$,
$\left|z\rangle\right. =
(\left|\uparrow\downarrow\rangle\right.+\left|\downarrow\uparrow\rangle\right.)/\sqrt{2}$.
The Hilbert space dimension is conserved by imposing the constraint
$s_i^\dagger s_i + \sum_\alpha  t^\dagger_{i\alpha} t_{i\alpha} = 1$ on every site $i$.
(In the presence of a Zeeman field, a rotated triplet basis with
$t^\dagger_{i+}=(t^\dagger_{ix}+it^\dagger_{iy})/\sqrt{2}$,
$t^\dagger_{i-}=(t^\dagger_{ix}-it^\dagger_{iy})/\sqrt{2}$,
$t^\dagger_{i0}=t^\dagger_{iz}$ is useful as well.)

The Heisenberg Hamiltonian (\ref{H}) can now be formulated in terms of the
bond operators $\{s_i, t_{i\alpha}\}$,
for details see Refs.~\onlinecite{bondop,kotov,MatsumotoNormandPRL}.
To treat the paramagnetic phase, the following re-interpretation of the
formalism is useful:\cite{kotov}
Starting from a background product state of singlets on all dimers,
$|\psi_0\rangle = \prod_i s_i^\dagger |0\rangle$,
the operators $t^\dagger_{i\alpha}$ can be viewed as creating local triplet
excitations in the singlet background.
(Formally, this is achieved by setting $s_i=s_i^\dagger=1$ which implies
$t^\dagger_{i\alpha}s_i \to t^\dagger_{i\alpha}$.)
The constraint then takes the form
$\sum_\alpha  t^\dagger_{i\alpha} t_{i\alpha} \leq 1$.
(So far, the procedure is exact.)

Upon expressing $\cal H$ \eqref{H} in bond operators,
products of two spin operators convert into terms with two, three, and four triplet operators,
${\cal H}={\cal H}_2+{\cal H}_3+{\cal H}_4$;
the biquadratic spin term in $\Hcoll$ \eqref{Hcoll} contains up to 8 triplets.
For the ideal bct structure and an external field in $z$ direction, the
bilinear part ${\cal H}_2$ reads:
\begin{equation}
\label{H2}
{\cal H}_2 =
\sum_{\vec{q}\alpha}\left\{\left(A_{\vec{q}}-\alpha H\right) t^\dagger_{\vec{q}\alpha}
t_{\vec{q}\alpha} +
\frac{B_{\vec{q}}}{2}\left(t_{\vec{q}\alpha}t_{-\vec{q}\bar{\alpha}}+h.c.\right) \right\}
\end{equation}
with $\alpha=+,0,-$, $\bar{\alpha} = -\alpha$,
$A_{\vec{q}} = J + B_{\vec{q}}$,
$B_{\vec{q}} = 2 J' \gamma_{{\vec q}\parallel} + 2 J_{2z} \gamma_{{\vec q}z}$ and
\begin{eqnarray}
\gamma_{{\vec q}\parallel} &=& (\cos q_x \!+\! \cos q_y)/2, \nonumber\\
\gamma_{{\vec q}z}         &=& \cos (q_x/2) \cos (q_y/2) \cos q_z.
\label{eq:def_gammas}
\end{eqnarray}
The coupling constant $\Jtz$ appearing in $A_{\vec{q}}$, $B_{\vec{q}}$ is
$\Jtz = \Jz^{11} + \Jz^{22} - \Jz^{12} - \Jz^{21}$.
The 3d momentum $\vec q$ runs over the Brillouin zone of the
bct lattice, spanned by the primitive translations
$\hat {\vec q}_1=(2\pi,0,-\pi)$, $\hat {\vec q}_2=(0,2\pi,-\pi)$, $\hat {\vec q}_3=(0,0,2\pi)$
in reciprocal space.

The harmonic (or linearized) approximation consists in treating only the bilinear
part ${\cal H}_2$ of the Hamiltonian; both the hard-core constraint and the higher-order terms
${\cal H}_3$, ${\cal H}_4$ (given in App.~\ref{kotovapp}) are neglected.
With the Bogoliubov transformation $t_{\vec{q}\alpha} =
u_{\vec{q}}\tau_{\vec{q}\alpha}+v_{\vec{q}}\tau^\dagger_{-\vec{q},\bar{\alpha}}$
the Hamiltonian ${\cal H}_2$ can be diagonalized,
with eigenvalues
\begin{equation}
\label{eq2}
\omega_{\vec{q}\alpha}  = \sqrt{A_{\vec{q}}^2-B_{\vec{q}}^2} -\alpha H
\end{equation}
and Bogoliubov coefficients
\begin{equation}
\label{eq3}
u_{\vec{q}}^2, v_{\vec{q}}^2 = \pm \frac{1}{2} +
\frac{A_{\vec{q}}}{2\omega_{\vec{q}0}}\,,~~
u_{\vec q} v_{\vec q} = - \frac{B_{\vec q}}{2\omega_{\vec q 0}}.
\end{equation}
In this linearized bond-operator theory, interactions between the order-parameter fluctuations --
represented by triplet quasiparticles -- are ignored.

For $|\Jtz|/J < 2$ the dispersion minimum of the magnetic modes is at
${\vec q}_\parallel = (\pi,\pi)$.
There, $\omega_{\vec{q}\alpha}$ is independent of $q_z$
due to frustration -- hence effectively two-dimensional.
Expanding near $(\pi,\pi)$ yields:
\begin{equation}
\omega_{\vec{q}0} =
\Delta + \frac{c^2}{\Delta} \left(k_\parallel^2 + \eta k_x k_y \cos q_z \right)
\end{equation}
where ${\vec k}_\parallel = {\vec q}_\parallel - (\pi,\pi)$ and
$\eta = -\Jtz/(2J')$.
The spin gap and velocity are given by
$\Delta=\sqrt{J(J-4J')}$ and $c = \sqrt{J J'/2}$.

For larger unit cells, multiple triplon operators are introduced, and
the Bogoliubov transformation needs to be performed numerically.

\subsection{Beyond the harmonic approximation}
\label{sec:kotov}

Triplon interactions effects are important to lift the degeneracy of the
dispersion along $(\pi,\pi,q_z)$.
The most important interaction correction
arises from the hard-core constraint, which is conveniently implemented using
an infinite on-site repulsion between the bosons:\cite{kotov}
\begin{equation}
\label{HU}
{\cal H}_U = U\sum_{i\alpha\beta}
t^\dagger_{i\alpha}t^\dagger_{i\beta}t_{i\alpha}t_{i\beta}, \quad
U\rightarrow\infty\; .
\end{equation}
As proposed by Kotov {\em et al.} \cite{kotov}, this hard-core term can be treated by a ladder
summation of scattering diagrams.
together with a self-consistent one-loop approximation for the
self energy, Fig.\,1 of Ref.~\onlinecite{kotov}.
This method is also known as Brueckner approach in particle theory,
the small parameter being the density of triplet bosons,
which at zero temperature is given by $\sum_{{\vec q}\alpha} v_{{\vec q}\alpha}^2$.

Here, we have employed the formalism of Ref.~\onlinecite{kotov}
at finite fields (App.~\ref{kotovapp}).
In the present problem, it is not sufficient to treat the quartic terms in ${\cal H}_4$
in a mean-field (Hartree-Fock) approximation:
processes of second order in ${\cal H}_4$ are needed to obtain the {\em leading}
contribution $\propto \Jz^4$ to the unfrustrated second-neighbor hopping in $z$ direction.
The calculations involve the self-consistent solution of integral equations
and are performed numerically on lattices with up to $16^3$ sites.

The approach of Ref.~\onlinecite{kotov} has been shown
to give results in good quantitative agreement with Quantum Monte Carlo
and series-expansion methods, regarding e.g. phase boundaries and
magnetic excitations of the bilayer Heisenberg model.
Although the Brueckner method contains a re-summation of an infinite series of
diagrams, it is not designed to capture critical behavior beyond mean-field.

\subsection{Ordered phases}

The bond-operator method can be generalized to magnetically ordered states
by taking into account the appropriate condensate.
Technically, an expansion is then performed around a symmetry-broken product
state $|\psi_0\rangle$ which replaces the singlet state.
As demonstrated by Sommer {\em et al.},\cite{sommer} a consistent
description of the fluctuations is obtained by applying a harmonic
approximation after a rotation of the basis vectors in the four-dimensional
Hilbert space of each dimer.

Here we adopt the formalism of Ref.~\onlinecite{sommer}, briefly
summarized in the following, to the bct lattice geometry.
The rotated basis operators, replacing $\{s_i^\dagger,t_{i\alpha}^\dagger\}$, are
\begin{eqnarray}
\tilde{s}^{\dagger}_i \!&=&\! \frac{1}{\sqrt{1+\lambda^2}}
  \Big[s^{\dagger}_{i}+\frac{\lambda \text{e}^{i\vec{Q}\vec{R_i}}}{\sqrt{1+\mu^2}}
  (t^{\dagger}_{ix}+i\mu t^{\dagger}_{iy})\Big], \nonumber \\
\tilde{t}^{\dagger}_{ix} \!&=&\! \frac{1}{\sqrt{1+\lambda^2}}
  \Big[-\lambda \text{e}^{i\vec{Q}\vec{R_i}}
  s^{\dagger}_{i}+\frac{1}{\sqrt{1+\mu^2}}\big(t^{\dagger}_{ix}+i\mu t^{\dagger}_{iy}\big)\Big], \nonumber \\
\tilde{t}^{\dagger}_{iy} \!&=&\! \frac{1}{\sqrt{1+\mu^2}}
  \left(t^{\dagger}_{iy}+i\mu t^{\dagger}_{ix}\right), \nonumber \\
\tilde{t}^{\dagger}_{iz} \!&=&\! t^{\dagger}_{iz},
\label{field-op-trafo}
\end{eqnarray}
where $\lambda$ and $\mu$ are condensate amplitudes.
The role of the singlet product state is now taken by
$|\psi_0\rangle = \prod_i \tilde{s}_i^\dagger |0\rangle$.
For $\lambda=\mu=0$ we have the original ``paramagnetic'' bond operators,
while $\lambda=1$, $\mu=0$ describes a classical N\'eel state $|\psi_0\rangle$
and its local excitations
(note that we have chosen the direction of the staggered magnetization to be in $x$ direction).
Finally, for $\mu=1$ and $\lambda\to\infty$ the product state is the
fully polarized state with all spins in $z$ direction.
Below, we will employ ordering wavevectors $\vec Q = (\pi,\pi,0)$ or $(\pi,\pi,\pi)$,
both describing states with ferromagnetic correlations between 2nd vertical neighbors
and reflecting the Z$_2$ degeneracy w.r.t. the relative orientation of neighboring
layers.

The Hamiltonian can be re-written in terms of the $\{\tilde{s}$,$\tilde{t_\alpha}\}$
operators; the corresponding lengthy expressions can be found in
Ref.~\onlinecite{sommer} and will not be reproduced here.
The condensate parameters $\lambda$ and $\mu$ are determined by minimizing
$\langle\psi_0|H|\psi_0\rangle$; this can be shown to be equivalent to
eliminating {\em linear} $\tilde{t}^\dagger$ terms in $\cal H$.
Then, as in the paramagnetic case,
the $\tilde{t}_\alpha^\dagger$ can be treated as excitations on top of
a background state, and  $\tilde s$ will be formally set to unity.
Subsequently, the Hamiltonian admits a harmonic approximation,
by only keeping the quadratic terms in $\tilde{t}_\alpha$.
The resulting ${\cal H}_2$ is solved by a Bogoliubov transformation,
leading to new quasiparticles $\tau_{{\vec q}\alpha}$.
For $\lambda=1$ and $\mu=0$, ${\cal H}_2$ is equivalent to
conventional linear spin-wave theory of a N\'eel-ordered antiferromagnet.

In summary, the modified bond-operator approach
interpolates between the triplon description of the
paramagnet, spin waves of the antiferromagnet, and the flipped-spin
quasiparticle physics of the field-polarized ferromagnet.
Within the harmonic approximation, the phase transitions
turn out to be of second order, and the ordered phases have the correct
number of Goldstone modes.
We also note that an extension beyond the harmonic level is not obvious:
Taking into account a hardcore repulsion of the $\tilde{t}_\alpha$
as above leads to a violation of the Goldstone theorem.
Therefore our quantitative calculations in the ordered phases below
will be restricted to the harmonic level.
This is expected to be a reasonable approximation away from
the phase transitions, and is also qualitatively correct near phase
transition above the upper-critical dimension.
However, we cannot capture critical behavior beyond mean-field,
including possible logarithmic corrections occuring at phase transitions
being at the upper-critical dimension (like a BEC transition in 2d).

\subsection{Observables}

In the harmonic approximation,
static observables like magnetizations are calculated as
expectation values with the ground state of ${\cal H}_2$
which is the vacuum of the Bogoliubov-transformed operators $\tau_{{\vec k}\alpha}$.

It may be tempting to calculate e.g. the $T=0$
uniform magnetization from the field dependence of the ground-state energy,
$M=-\partial E_0/\partial H$.
However, this procedure is {\em incorrect} in magnetically ordered phases
for the following reason:
In an ordered phase, the condensate parameters and hence the basis vectors
are in general field-dependent, i.e., the expansion is done around a
field-dependent product state.
This implies that the quadratic part of the Hamiltonian will be field-dependent.
In other words, in the harmonic approximation ${\cal H}_2$ corresponds to
a {\em different} Hamiltonian for each field.
Then, $-\partial E_0/\partial H$ contains, apart from the magnetization,
an additional contribution arising from the field dependence of ${\cal H}_2$.
We note that $M=-\partial E_0/\partial H$ is sometimes used in spin-wave
theories for canted antiferromagnets -- there, it yields incorrect results as well.
However, the deviations from $\langle S_z \rangle$ are often small.

Dynamic properties, like the cross section for inelastic neutron scattering,
can be expressed in terms of the Green's functions of the triplon quasiparticles.
Beyond the harmonic approximation, this route needs to be taken for static observables
as well.

\subsection{Relation to single-boson description}

Let us close this section with comments on the relation
between the advocated bond-boson approach and effective theories
based on a single-boson description often used
in the field-driven cases.\cite{giamarchi}

It is apparent, that near $H_{c1}$ only the $\tau_+$ boson is low in energy
(equivalent to the $\Phi = \phi_x+i\phi_y$ mode of Sec.~\ref{sec:sym}),
hence $\tau_+$ is the only required degree of freedom in a critical theory
for the $H_{c1}$ transition.
There are, however, a few subtleties:
(i) Scattering processes between low-energy and high-energy bosons
contribute to terms in the low-energy theory.
This is in fact the case in the ideal bct lattice when $\Jzz=0$ in
the Hamiltonian. Then, the bare boson dispersion is independent of $q_z$ at
wavevectors $(\pi,\pi,q_z)$, and interactions are required to lift this degeneracy.
As scattering processes between $\tau_+$ bosons conserve particle number
and hence are absent at $T=0$, one may conclude that the dispersion remains flat.
This is incorrect, because scattering processes with the high-energy particles $\tau_{-,0}$
are no longer number-conserving, inducing a finite dispersion.
(ii) To fully describe the condensate with canted order throughout the phase diagram,
a single complex parameter is insufficient, as clearly seen from the full
bond-operator description (where both $\lambda$ and $\mu$ are needed).
Using a condensate of $\tau_+$ only is appropriate in the limit $J' \ll J$,
but this cannot describe states close to a quasiclassical N\'eel state.

This concludes our description of the methods.


\section{Case (A): Equivalent layers and full frustration}
\label{sec:eq}

This section discusses in detail the situation (A) with an ideal bct structure of dimers,
while situation (B)  -- a structure with inequivalent layers and two dimers per unit
cell -- will be studied in Sec.~\ref{sec:ineq}.
For the purpose of comparison with experiments, most numerical results shown below
are for parameter values of possible relevance to the material \bacu.

The dimensional reduction observed in \bacu, i.e., the 2d value of the
shift exponent describing the BEC phase boundary,
$T_c \propto (H-H_{c1})^\psi$ with $\psi=1$,
was originally discussed in terms of frustration only,
i.e. scenario (A).\cite{sebastian}
While the neutron scattering results of Ref.~\onlinecite{rueggbacusio}
point to a somewhat different origin of quasi-2d behavior,
namely scenario (B) with inequivalent dimers,
the physics of the ideal bct structure is a very interesting and instructive
situation to study.
Some of the results described below were reported by us in Ref.~\onlinecite{ORMV}.

\subsection{Paramagnetic phase: Dispersions}
\label{sec:eqpara}

For one dimer per unit cell, there will be a single branch
of magnetic triplet excitations in the zero-field singlet phase.
Its in-plane dispersion caused by $J'$ is unfrustrated, but
the $c$ axis dispersion is heavily influenced by frustration:
Away from the in-plane dispersion minimum at ${\vec q}_\parallel=(\pi,\pi)$
it is dominated by the frustrated
hopping between adjacent layers, which leads to a (bare) dispersion of the form
$\Jz \cos(q_x/2) \cos(q_y/2) \cos q_z$.
This vanishes at ${\vec q}_\parallel=(\pi,\pi)$, and only second-neighbor
hopping contributes, with an additional dispersion $\Jzz \cos 2 q_z$.

A key question, also relevant for the critical behavior,
is the fate of the vertical dispersion in a model with bare $\Jzz=0$.
The arguments in Refs.~\onlinecite{sebastian} suggested that the dispersion
would then be $q_z$-independent along $(\pi,\pi,q_z)$,
rendering the critical behavior at $H_{c1}$ truly two-dimensional.
However, interaction processes invariably
induce a symmetry-allowed effective vertical second-neighbor hopping\cite{ORMV}
of order $J_z^4/J^3$ -- this effect is the one responsible
for the absence of true dimensional reduction
(with the exception of the high-field situation $H\geq H_{c2}$,
see Sec.~\ref{sec:hc2} below).

\begin{figure}[t]
\epsfxsize=3.2in
\centerline{\epsffile{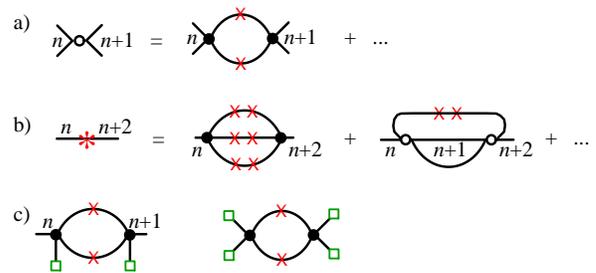}}
\caption{
Diagrams occurring in the perturbation expansion of the
order-parameter theory ${\cal S}_\phi$ \eqref{sphi}.
The solid lines are $\phi$ propagators
(depending on in-plane momentum $\vec k_\parallel$ and layer index $n$),
the full circle is the local four-point vertex ($\propto u_0$),
the cross is the frustrated inter-layer hopping ($\propto \eta k_x k_y$).
a) Inter-layer density interaction [open circle, this includes both $u_1$ and $u_2$ terms in
${\cal S}_{AB}$ \eqref{sab}],
generated from a $u_0^2$ process.
For negative $u_1$, the $u_1 (\vec\phi_n \cdot \vec\phi_{n+1})^2$ term leads to collinear
spin correlations in $z$ direction.
b) Unfrustrated second-neighbor vertical hopping (star, $\propto \eta'$),
generated from interaction processes.
This hopping is responsible for 3d behavior at lowest energies, irrespective
of the inter-layer frustration.
c) Additional diagrams present in the ordered phase;
the open square denotes the coupling to the condensate.
The first diagram corresponds to nearest-neighbor vertical hopping,
the second is the leading ``vertical'' contribution to the free energy.
}
\label{fig:dgr1}
\end{figure}

\begin{figure}[t]
\epsfxsize=2.4in
\centerline{\epsffile{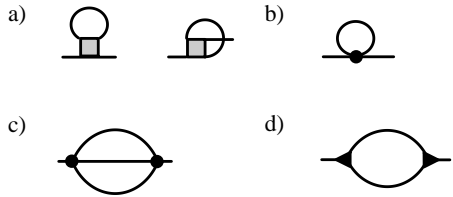}}
\caption{
Diagrams in the Brueckner bond-operator approach.
The solid lines are $t_{{\vec q}\alpha}$ propagators,
here depending on the full 3d momentum $\vec q$.
a) Self-energy from the hard-core repulsion; the shaded square
is the effective four-point vertex $\Gamma$ \eqref{Gamma}
obtained from a ladder summation.
b) Leading self-energy from the quartic term ${\cal H}_4$,
with the circle corresponding to $J_4$.
c) Second-order self-energy in ${\cal H}_{4}$ --
its ${\cal H}_{4z}^2$ portion is required to obtain
the leading term in the vertical dispersion on the ideal bct lattice.
d) Second-order self-energy in ${\cal H}_{3z}$, where the triangle is the
three-point vertex of strength $J_3^\pm$.
The momentum space structure of the vertex suppresses this diagram at ${\vec q}_\parallel =
(\pi,\pi)$.
}
\label{fig:dgr2}
\end{figure}

Let us briefly repeat the perturbative arguments for the vertical dispersion.
Within the effective order-parameter theory for $\vec\phi$,
we need to look for processes which generate a contribution to the $\eta'$
term in ${\cal S}_A$, Eq.~\eqref{phi4},
from the inter-layer interactions ${\cal S}_{AB}$, Eq.~\eqref{sab}.
The leading diagrams are in Fig.~\ref{fig:dgr1}b, and are
$\propto u_0^2 \eta^6$ and $\propto u_{1,2}^2 \eta^2$.
The microscopic identification $\eta\propto\Jz$, $u_{1,2}\propto\Jz^2$
suggests that both diagrams are of order $\Jz^6$.
Higher-order processes will not change this result
(except, perhaps, at a critical point, see Sec.~\ref{sec:eqppd} below).
However, the order parameter theory misses interactions between
staggered and uniform magnetization fluctuations, as the latter are not
contained in ${\cal S}_\phi$.
This becomes clear when discussing the same physics in the bond-operator language.
Relevant interaction terms are the hard-core term ${\cal H}_U$ \eqref{HU} and
the three- and four-point vertices arising from $\Jz$, ${\cal H}_{3z}$ \eqref{h3z},
and ${\cal H}_{4z}$ \eqref{h4z}.
(Note that the physics of ${\cal H}_{3z}$ and ${\cal H}_{4z}$ is absent from
the effective order-parameter theory.)
We will ignore effects of the in-plane part of ${\cal H}_{4}$ beyond the Hartree contribution,
because the dominant overall renormalization arises from ${\cal H}_U$,\cite{kotov}
and qualitative changes from ${\cal H}_{4\parallel}$ are not expected.
Taking into account ${\cal H}_U$ yields a dispersion along $(\pi,\pi,q_z)$
proportional to $\Jz^6 \cos (2 q_z)$. This is easily understood, as the self-consistent
Hartree diagram, Fig.~\ref{fig:dgr2}a, contains the process in Fig.~\ref{fig:dgr1}b2.
Now consider the ${\cal H}_{4z}^2$ portion of the diagram in Fig.~\ref{fig:dgr2}c.
Remarkably, it does generate a vertical dispersion proportional to $\Jz^4$
(see App.~\ref{kotovapp}) -- hence
this the {\em leading} contribution to the interaction-generated vertical dispersion.
We have therefore incorporated this diagram into the Brueckner approach.
Finally, there is ${\cal H}_{3z}$.
Fig.~\ref{fig:dgr2}d shows the leading self-energy diagram, which, however, turns
out to vanish at $\vec q_\parallel = (\pi,\pi)$.
The contribution from ${\cal H}_{3z}$ to the vertical dispersion
is of order $\Jz^6$; hence we will ignore ${\cal H}_{3z}$ altogether.

\begin{figure}[t]
\epsfxsize=3.4in
\centerline{\epsffile{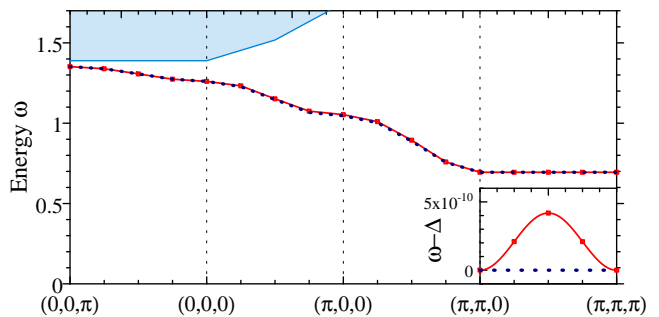}}
\caption{
Triplon dispersion (solid) and boundary of the two-particle continuum (shaded),
calculated using the Brueckner bond-operator approach
(Sec.~\ref{sec:kotov} and App.~\ref{kotovapp})
on an $8^3$ lattice.
The parameter values are $J=1$, $J'/J=0.15$, $-\Jtz=\Jfz=0.2J'$, $J_{3z}=0$, $H=0$.
The inset shows an energy zoom into the almost flat vertical dispersion.
For comparison, the dotted line shows a dispersion calculated within the
harmonic approximation, with parameter values chosen to {\em match} the
Brueckner results at wavevectors $(0,0,0)$, $(\pi,\pi,0)$, and $(0,0,\pi)$:
$J=1.047$, $J'/J=0.140$, $-\Jtz/J'=0.196$.
}
\label{fig:match}
\end{figure}

A full result for the dispersion is shown in Fig.~\ref{fig:match}.
The parameter values are chosen with an eye towards \bacu, i.e.,
we have tried to match the in-plane dispersion averaged over the modes
observed in Ref.~\onlinecite{rueggbacusio}, i.e., $J'/J=0.15$.
The vertical coupling is somewhat arbitrarily chosen as $-\Jtz=\Jfz=0.2J'$,
giving a dispersion at the unfrustrated point, i.e., along $(0,0,q_z)$,
of $(-0.096 J \cos q_z)$.
The dispersion at the frustrated point is tiny,
roughly $(-2 \cdot 10^{-10} J \cos 2 q_z)$,
its bandwidth scales as $\Jz^4$ as expected.

In Fig.~\ref{fig:match} we illustrate that the in-plane dispersion
obtained from the Brueckner approach can be reproduced using the
harmonic bond-operator approximation describing non-interacting triplons
with {\em renormalized} parameters.
(Similar observations were made earlier e.g. in the context of spin ladders.\cite{eder98})
The leading renormalization is in $J$: triplon repulsion pushes the dispersion to higher
energies, hence the ``harmonic'' $J$ is larger than the ``true'' (Brueckner) $J$.
Of course, the renormalization in general depends on temperature,
magnetic field and other parameters.
Using the zero-field renormalized parameters gives a reasonable account of the
field dependence of the spin gap, and overestimates $H_{c1}$ only by a few percent
(compared to the Brueckner approach).
In the ordered phases and in the more complicated case of inequivalent
layers (Sec.~\ref{sec:ineq}), where a Brueckner calculation is no longer feasible,
we will exploit this fact, i.e., we will work with the harmonic approximation
and parameter values chosen to match experimental data.

\begin{figure}[b]
\epsfxsize=3.4in
\centerline{\epsffile{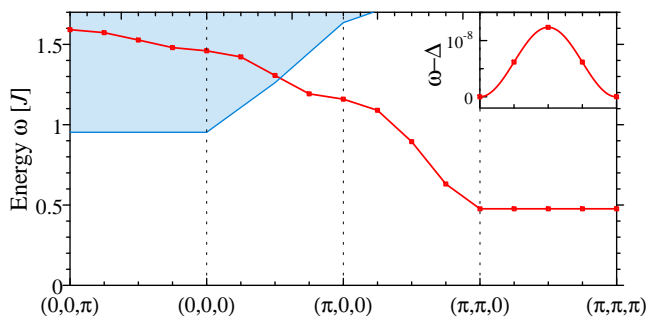}}
\caption{
Triplon dispersion (solid) and two-triplon continuum (shaded)
as in Fig.~\ref{fig:match}, but with parameter values closer to
the zero-field transition: $J=1$, $J'/J=0.25$, $-\Jtz=\Jfz=0.2J'$, $J_{3z}=0$, $H=0$.
Compared to Fig.~\ref{fig:match}, the bandwidth along $(\pi,\pi,q_z)$ is
significantly enhanced.
}
\label{fig:jp025}
\end{figure}

\begin{figure}[t]
\epsfxsize=3.4in
\centerline{\epsffile{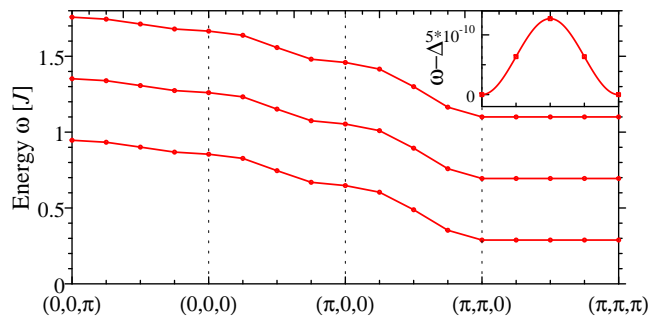}}
\caption{
Triplon dispersion as in Fig.~\ref{fig:match},
but now in a finite field: $J=1$, $J'/J=0.15$, $-\Jtz=\Jfz=0.2J'$, $J_{3z}=0$, $H=0.4$.
}
\label{fig:h04}
\end{figure}

Moving closer to the zero-field transition by increasing $J'/J$,
the overall triplon bandwidth increases, see Fig.~\ref{fig:jp025}.
The interaction-generated bandwidth $E_z$ along $(\pi,\pi,q_z)$ strongly increases,
as the relevant energy denominator in the fourth-order
expression is given by the third power of an averaged triplon energy.
The induced bandwidth will also depend on temperature,
but this effect is exponentially suppressed for $T<\Delta$.
From the numerical results, we are not able to track $E_z$
close to the phase transition, due to discretization errors on the
finite lattice.
Finally, upon applying a field, the triplon modes simply split as expected,
Fig.~\ref{fig:h04}.

After having established the vertical {\em second}-neighbor coupling,
it is worth discussing the {\em first}-neighbor correlations.
In the field-theory language, these arise from the $u_1$ term in ${\cal S}_{AB}$ \eqref{sab}
and are of orthogonal (collinear) type for $u_1>0$ ($u_1<0$),
as the $u_1$ term is insensitive to the sign of $\vec\phi_{A,B}$.
These correlations can be seen in four-point spin correlators only.
The peculiar momentum-space structure of the $\eta$ term
implies that two-point spin correlators between even and odd planes
are {\em strictly zero} within the order-parameter theory in its paramagnetic phase.
(The microscopic bond-operator approach shows weak ferromagnetic correlations
between neighboring planes, which do not become long-ranged near the ordering
transition.)

Microscopically, the $u_1$ term may originate from a biquadratic spin coupling as in
$\Hcoll$ \eqref{Hcoll}. As above, one has to ask about the value of $u_1$
in the absence of a bare $\Jcoll$.
In fact, $u_{1,2}$ will be interaction-generated as well, with the leading
diagram $\propto \Jz^2$ shown in Fig.~\ref{fig:dgr1}a.
Care needs to be taken when contracting two $u_0$ vertices
because of the vector structure of the interactions.
A straightforward analysis shows $u_1,u_2<0$ with $u_2/u_1 = (N+4)/4$,
where $N$ is the number of order parameter components.
$u_1<0$ means that collinear correlations are favored -- this is consistent
with the quasiclassical order-from-disorder mechanism operative inside
the ordered phases, see Sec.~\ref{sec:eqofd}.
(Note that $u_2$ acquires additional contributions $\propto J_z$ near
$H_{c1,2}$ from a mean-field-like interaction of {\em uniform} magnetizations.)

\subsection{Antiferromagnetic phase: Order from disorder}
\label{sec:eqofd}

We now turn to the physics inside the antiferromagnetically ordered phases;
the discussion of the phase transitions will be postponed to
the following subsections.

As mentioned in the introduction, for classical moments on the bct lattice,
antiferromagnetic layers are completely decoupled (provided $\Jzz=0$).

\subsubsection{Spin-wave theory}

The effect of quantum fluctuations on the frustrated bct antiferromagnet has been
initially studied using semiclassical spin-wave theory,\cite{rastelli1,rastelli2,shender}
and we briefly summarize the results here.

The simplest calculation, for a model with spins $S$ on sites of a bct lattice,
assumes a spiral order with wavevector $(\pi,\pi,Q_z)$.
In linear spin-wave approximation, the spin-wave spectrum is degenerate
along the line $(\pi,\pi,q_z)$, signaling frustration.
However, the ground state energy contains an inter-layer contribution
of the form $(-\Jz^2 S/J' \cos^2 Q_z)$, thus favoring {\em collinear} order
with $Q_z = 0$ or $Q_z=\pi$ --
this term arises from the zero-point energy of high-energy spin-wave modes.
Including spin-wave interactions to order $1/S$ removes the degeneracy
in the dispersion along $(\pi,\pi,q_z)$, consistent with the assumed ordering
wavevector -- the dispersion is proportional to $\sqrt{S} |\Jz \sin (q_z/2)|$, i.e.,
corresponds to vertical first-neighbor hopping.\cite{rastelli2}
Thus, this calculation predicts ferromagnetic order between second-neighbor
planes, and collinear order between neighboring planes.
A subsequent, more detailed, calculation\cite{shender} relaxed the assumption
of a single ordering wavevector and analyzed more general ordered states to
high orders in $1/S$.
As a result, for $S=1/2$ a state with {\em antiferromagnetic}
order between second-neighbor planes and collinear order between neighboring
planes was found be selected by energy contributions of order
$\Jz^4/(JS)^3$ and $\Jz^6/(J^5 S)$.
Thus, fully 3d order is stabilized within the stacks of even and odd planes.
The residual Z$_2$ degeneracy specifying the relative orientation of
the ``even'' and ``odd'' order parameter (Sec.~\ref{sec:z2op})
is left intact and hence spontaneously broken in the ordered state.

We note that the tendency towards collinear correlations can be mimicked\cite{coleman}
by a biquadratic term in the Hamiltonian of the form \eqref{Hcoll},
with $\Jcoll \propto -\Jz^2/J'$ -- we will exploit this in the triplon
bound-state calculation.

One may also consider order-from-disorder on the bct lattice in the different
situation with non-vanishing second-neighbor vertical coupling $\Jzz$.
Then, 3d order in each of the tetragonal subsystems is already established at
the classical level, but the two order parameters are decoupled.
Quantum fluctuations are only needed to produce the collinear coupling ($\propto
\Jz^2/J'$), reducing the degeneracy to the residual Z$_2$ symmetry.
(This latter order-from-disorder mechanism is similar to the one
in the much-studied $J_1-J_2$ Heisenberg model on the square lattice,
in the limit of large $J_2$.)

\subsubsection{Field theory}

It is instructive to re-phrase the order-from-disorder mechanism in terms
of the order-parameter field theory of Sec.~\ref{sec:sym},
with the relevant diagrams shown in Fig.~\ref{fig:dgr1}.
First, the magnetization orientations between adjacent layers
are determined by the $u_1 (\vec\phi_n \cdot \vec\phi_{n+1})^2$ term.
Indeed, we have argued above for $u_1<0$ which favors collinear correlations.
Second, the type of 3d ordering of second-neighbor layers is
determined by the $\eta' \vec\phi_n \cdot \vec\phi_{n+2}$ term.
Third, the coupling to the condensate allows an effective
nearest-neighbor hopping in $z$ direction, which was symmetry-forbidden in
the paramagnetic phase.
The diagrams, Fig.~\ref{fig:dgr1}c, show that the order-from-disorder contributions
to the vertical dispersion and to the free energy
scale with the square and the fourth power of the order parameter, respectively.

\subsubsection{Bond operators}

In the Brueckner bond-operator approach, we can follow the properties in the
paramagnetic phase up to the phase transition.
From the single-particle dispersion, Fig.~\ref{fig:match}, we can read off $\eta'$;
we find $\eta'<0$, i.e., a dispersion minimum at $(\pi,\pi,0)$ and $(\pi,\pi,\pi)$.
Triplon condensation at this wavevector implies ferromagnetic orientation
between second-neighbor layers.
This may appear inconsistent with the above-mentioned result of
Ref.~\onlinecite{shender},
but the latter strictly only applies in the semiclassical limit.
The collinear inter-layer correlations are contained in four-point triplon correlators.
In those cases where we find a triplon bound state (Sec. \ref{sec:bound} below),
its internal structure implies collinear correlations and
its dispersion has a minimum at wavevector $(0,0,0)$.
Condensation of this bound state implies uniform bond order (see Sec.~\ref{sec:z2op})
and again ferromagnetic orientation between second-neighbor layers.
[The ordering pattern of Ref.~\onlinecite{shender} would correspond to
condensation of $\Psi$ at $(0,0,\pi)$.]

Inside the ordered phases, our bond-operator calculations
are restricted to the harmonic approximation.
Here, the dispersion degeneracy along $(\pi,\pi,q_z$) is not lifted:
there are too many Goldstone modes.
Nevertheless, we expect energy-integrated properties like magnetizations
to be semi-quantitatively correct.
Some results in the field-induced canted phase are shown in
Fig.~\ref{fig:mag11} below.

\subsection{Zero-field quantum phase transition}
\label{sec:eqppd}

The quantum paramagnet can be driven into an antiferromagnetically
ordered state at zero field by increasing $J'/J$.
Experimentally, the application of pressure changes $J$ and $J'$ via a
modification of bond lengths and angles.
The coupled-dimer material \tcc\ is driven into an ordered state
upon application of pressure, hence $J'/J$ is increased here;
for \bacu\ systematic pressure studies have not been performed to our knowledge.

In an unfrustrated system, the zero-field ordering transition
breaks the SU(2) spin symmetry and has dynamic exponent $z=1$.
In a spatially anisotropic, i.e., layered, system with a 3d scale $E_z$,
the quantum critical behavior is 2d for energies or temperatures above $E_z$
while it is 3d below $E_z$.
Importantly, $E_z$ is essentially given by the bandwidth of the vertical dispersion
near the ordering wavevector.

For the frustrated bct system, several complications arise
which we briefly address in the following:
There is the additional Z$_2$ symmetry to be broken,
and the vertical dispersion consists of a bare frustrated part
and an {\em interaction-generated} unfrustrated part.

\subsubsection{Interaction-generated dispersion}

If the bare $\Jzz$ vanishes, then a vertical dispersion along $(\pi,\pi,q_z)$
(determining the 3d crossover scale)
only arises from interactions.
The leading term is of second order in a four-point vertex (Sec.~\ref{sec:eqpara}),
it strongly depends on the distance to criticality because the
spin gap enters the denominator of the perturbative expression.

The simplest perturbative estimate is reliable in the paramagnetic phase for
not too large $\Jz$, but may fail at criticality due to a non-trivial flow of the
four-point vertex. This happens at or below the upper critical dimension $d_c^+$
and is signaled by infrared divergencies in the next-order diagrams.
The renormalization group thus has a rather interesting structure
(assuming a vanishing bare $\eta'$):
one-loop accuracy is sufficient to study the flow of the four-point vertex,
but the strongly relevant two-point vertex $\eta'$ is only generated at two loops.
(A similar situation has recently been studied in the context of spin chains
with a frustrated inter-chain coupling.\cite{starykh})
Here we shall not perform a consistent two-loop renormalization group treatment;
we expect that the scaling $E_z \propto \Jz^4$ continues to hold for small $\Jz$,
albeit with non-trivial finite-temperature corrections to the prefactor
in the quantum critical regime.
(The $\Jz^4$ scale is expected to evolve continuously into the
$\Jz^4$ scale of the ordered phase, Sec.~\ref{sec:eqofd},
which decides between ferro- and antiferromagnetic ordering between
second-neighbor layers.)

We also note that the bond-operator calculation partially accounts for the above
renormalization effects due to the self-consistent resummation of vertex diagrams.
(For the field-driven case, such a strong renormalization of $\eta'$ does not arise,
as the interaction processes necessarily involve the high-energy triplon branches,
due to particle-number conservation within the low-energy branch.)

\subsubsection{Bound state condensation and split transition}
\label{sec:bound}

As discussed in Sec.~\ref{sec:z2op}, the Z$_2$ symmetry can
be broken via condensation of the bond order parameter $\Psi$ before
magnetic order occurs.
$\Psi$ corresponds to a singlet bound state of two triplons.\cite{kotovbound}
While the transition to bond order may be studied using a mean-field approach,
the bound-state dynamics requires more effort.
As in Ref.~\onlinecite{kotovbound}, we have solved a Bethe-Salpeter equation
for two-triplon bound states.
The required input encompasses the single-triplon dispersion -- to be taken
from the Brueckner bond-operator calculation -- and the triplon four-point
scattering vertex.
As self-consistency in the two-particle sector is beyond reach,
it is reasonable to work with a bare scattering vertex,
with contributions from ${\cal H}_4$ and ${\cal H}_U$.
However, from our analysis it is clear that the effective biquadratic
inter-layer interaction,
represented by $u_1 (\vec\phi_n \cdot \vec\phi_{n+1})^2$ in the field theory,
is relevant for triplon attraction.
We therefore include by hand $\Hcoll$ \eqref{Hcoll} which mimicks the effect
of the collinear interaction;
details are relegated to App.~\ref{bseapp}.

Interestingly, for triplon attraction the couplings $(-\Jcoll)$ and $(+\Jfz)$
act in a very similar way: both can cause inter-layer binding.
While the triplon binding effect of $\Hcoll$ is easily understood in the
order-parameter language, this is more subtle with ${\cal H}_{4z}$.
First, we observe that the relevant part of ${\cal H}_{4z}$
involves the {\em uniform} (as opposed to staggered) degrees of freedom
of the dimers.
Second, $\Psi$ condensation represents bond order arising from quantum-mechanical
singlet formation, which is common for frustrated spin-1/2 systems.\cite{dimer}
(Note that the attractive force is {\em linear} in $\Jfz$.)

In zero field, we find a singlet bound state
below the two-particle continuum only for sufficiently strong attraction $(\Jfz-\Jcoll)$.
This is plausible, as small attraction in 2d causes only an exponentially shallow bound state,
while in 3d a finite attraction is needed for binding --
this simply follows from the properties of one particle with quadratic dispersion
moving in a potential well.
The wavefunction of the lowest bound state changes sign under 90-degree
in-plane rotations of the internal coordinate;
from the symmetry considerations in Sec.~\ref{sec:sym} we conclude
that this bound state corresponds to a $\Psi$ quantum as anticipated.
The bound-state dispersion has its minimum at $(0,0,0)$ and
a bandwidth along $(0,0,q_z)$ of $\Jz^4$, inherited from the two-triplon
continuum.
A condensation of this singlet bound state corresponds to the Z$_2$ symmetry
breaking advocated in Sec.~\ref{sec:z2op}.
The bound state is unaffected by a Zeeman field, hence for fields larger than
the triplon binding energy the bound state ceases to exist below the two-particle
gap.

\begin{figure}[t]
\epsfxsize=3in
\centerline{\epsffile{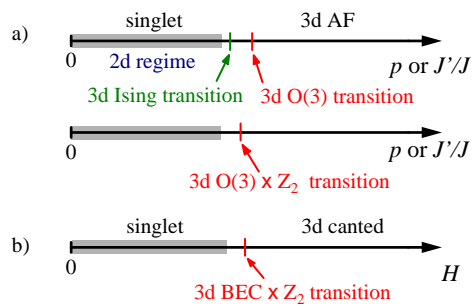}}
\caption{
Schematic $T=0$ phase diagrams for the bct lattice coupled-dimer model,
for the situation of equivalent layers with perfect frustration.
a) Pressure tuning - here generically two scenarios are possible
(where the second one applies for the parameters relevant to \bacu).
b) Field tuning.
}
\label{fig:schemeq}
\end{figure}

In conclusion, both of the following scenarios may be realized
(Fig.~\ref{fig:schemeq}):
(i) Two transitions:
Upon increasing $J'/J$, there is first a quantum Ising transition
with $z=1$ where bond order is established through condensation of $\Psi$.
This transition is asymptotically 3d:
the vertical $\Psi$ dispersion is of order $\Jz^4$ along $(0,0,q_z)$.
Subsequently, magnetic order will be established in a second transition
where $\vec\phi$ condenses and the SU(2) symmetry is broken.
The second transition has $z=1$ and is of conventional O(3) type,
in the sense that frustration is removed by the $\Psi$ condensate,
i.e., by the term $\lambda\langle\Psi\rangle {\vec \phi}_n \cdot {\vec \phi}_{n+1}$
[see Eq.~\eqref{phipsi}].
The distance of the transitions is determined by the triplon binding energy.
(ii) One transition:
the condensation of $\vec\phi$ breaks the SU(2)$\times$Z$_2$ symmetry in one step.
This transition is again asymptotically 3d;
a detailed study of the critical behavior shall not be undertaken here.

For the parameter values of possible relevance to \bacu, we see no indication
for bound states; hence case (ii) applies.
Unfortunately, the numerical Bethe-Salpeter calculation suffers from severe
finite-size effects, therefore we cannot reliably resolve small binding energies.
We also note that the inclusion of lattice effects can modify the behavior:
for instance, bond order could occur far before magnetic order sets in --
this applies not only to $T=0$ but also to the finite-$T$ transitions.

\subsubsection{Anisotropies and further perturbations}

Perturbations beyond the model Hamiltonian \eqref{H} can modify the asymptotic
critical behavior:
Those include magnetic anisotropies of Dshyaloshinski-Moriya or dipolar type,
as well as coupling to nuclear spins.
In particular, anisotropy terms are relevant perturbations to the O(3) critical points,
rendering the asymptotic critical behavior Ising-like.

\subsection{Field-driven quantum phase transitions}
\label{sec:eqhpd}

\begin{figure}[t]
\epsfxsize=3.3in
\centerline{\epsffile{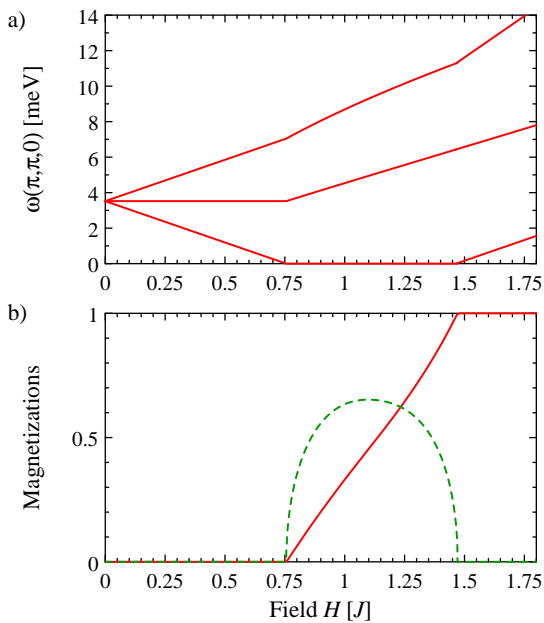}}
\caption{
a) Triplon energy gaps and
b) uniform (solid) as well as staggered (dashed) magnetization,
calculated as function of the external field $H$ using bond operators
in the harmonic approximation.
Parameter values describe the ideal bct lattice:
$J=4.66$ meV, $J'=0.5$ meV, $-\Jtz=0.1$ meV.
}
\label{fig:mag11}
\end{figure}

\subsubsection{Lower critical field $H_{c1}$}

The ground state of the quantum paramagnet is unaffected by applying a Zeeman field,
however, the triplon excitations will split, leading to a spin gap decreasing with
applied field.
At at critical field $H_{c1}$, given by the zero-field gap,
the lowest triplon mode becomes soft, and antiferromagnetic order
perpendicular to the field is established.
The phase transition is in the dilute Bose gas universality class,
with dynamical exponent $z=2$, and breaks a U(1) symmetry.\cite{ssbook}

As above, the energy scale $E_z$ for the dimensional crossover between 2d and 3d
is given by the vertical bandwidth of the single-particle dispersion
near the minimum wavevector.
In bct lattice coupled-dimer model with perfect inter-layer frustration,
this scale is again given by the sum of the bare vertical second-neighbor hopping,
$\Jzz$, and the corresponding interaction-induced contribution -- for the
latter the perturbative estimate $\Jz^4/J^3$ is now reliable even at the
transition, because the relevant $T=0$ interaction processes involve
high-energy triplons.
At the transition, U(1) and Ising symmetries are
broken simultaneously, Fig.~\ref{fig:schemeq}b,
as singlet bound states are irrelevant in finite field.
The phase transition is still of the dilute Bose gas type,
i.e. with $z=2$ and mean-field exponents, supplemented by logarithmic
corrections in the 2d regime.
(Magnetic anisotropies beyond the model \eqref{H} can modify the critical behavior at lowest
energies as noted above; a Dshyaloshinski-Moriya interaction may even smear out the
phase transition.)

For numerical calculations we resort to the harmonic bond-operator method.
To this end, we employ parameters
$J=4.66$ meV, $J'=0.5$ meV, $-\Jtz=0.1$ meV,
which roughly reproduce (within the harmonic approximation)
the \bacu\ mode dispersions (averaged over the multiple modes).
Results for mode gaps and magnetizations are shown
in Fig.~\ref{fig:mag11} (here $\vec Q=(\pi,\pi,0)$).


\subsubsection{Phase boundary: 3d critical behavior at elevated temperature?}

While inter-layer frustration leads to well-defined 2d quantum critical behavior
at intermediate energies or temperatures,
one can ask whether 3d critical behavior (with associated exponents)
is restored at higher energies --
this was proposed on the basis of the experiments on \bacu\ in
Ref.~\onlinecite{bacusio}.

While there is a sizeable vertical dispersion of the magnetic modes
away from in-plane wavevector $(\pi,\pi)$, which may be associated with 3d behavior,
the thermodynamics is (to leading order) determined by the density of states
of all magnetic modes.
Here the alternating sign of the inter-layer hopping (as function of in-plane momentum)
turns out to be crucial, because the 3d-like contributions to the density of states tend to cancel.
As a result, the density of states does never cross over to the power law
characteristic for 3d -- this applies to both zero-field and finite-field cases.

Turning to the location of the boundary of the ordered phase:
this can be estimated within the
order-parameter field theory by calculating the finite-temperature mass
correction, i.e., the temperature dependence of the order parameter self-energy,
on the paramagnetic side.
For a transition above the upper critical dimension, the lowest-order
estimate given by the Hartree diagram is usually sufficient;
what enters the Hartree diagram is precisely the density of states of the
magnetic modes.
Performing the integral numerically nicely shows a crossover from
$\Sigma(T)-\Sigma(0) \propto T^{3/2}$ for $T$ below the tiny $E_z$ to
$\Sigma(T)-\Sigma(0) \propto T$ above $E_z$,
but no further crossover to another well-defined power law.

This strongly suggests that experimentally observed deviations from the 2d critical power laws
(above 1\,K)
have nothing to do with 3d critical behavior, but instead indicate that one leaves the
critical regime (in the sense that the correlation length is no longer large,
or that the density of triplons in the ground state is no longer dilute).

\subsubsection{Upper critical field $H_{c2}$}
\label{sec:hc2}

For large external fields, $H>H_{c2}$, the ground state of the coupled-dimer
model is fully polarized, and a quantum phase transition to
a canted state occurs at $H_{c2}$ which is in the dilute Bose gas
universality class as well.

In contrast to the paramagnetic low-field phase,
the wavefunctions for the ground state and the one-particle excitations
above $H_{c2}$ are exactly known:
A Bloch wave of one spin flip on top of the ferromagnetic background
is an exact eigenstate of $\cal H$.
This has interesting consequences: although these spin-flip particles
have a hard-core interaction, scattering processes are rare
at low $T$, as the equilibrium particle density vanishes as $T\to 0$.
Thus, the interaction corrections to the bare dispersion are
exponentially suppressed above $H_{c2}$.\cite{batista}

Hence, in a model without bare $\Jzz$, the vertical mode dispersion
vanishes identically at $T=0$ for $H>H_{c2}$, and
will only be induced by thermal processes in the quantum
critical regime.
This suggests that the asymptotic critical behavior is two-dimensional.\cite{batista}

\subsection{Classical phase transition and N\'eel temperature}
\label{sec:tn}

Given the order-from-disorder mechanism, which generates
an effective 3d coupling proportional to the square of the
order parameter itself, and given the Mermin-Wagner theorem,
the obvious question about the nature of the finite-temperature
transition arises.
The issue appears particularly relevant for O(3) symmetry, i.e.,
in the zero-field case:
Starting in the ordered phase at low $T$, the system has a robust
(effective) vertical coupling of order $\Jz$,
and one would hence predict a N\'eel temperature $\TN$ roughly given by $J'/\ln (J'/\Jz)$
(in the limit of small $\Jz$, and $J'$ is the relevant in-plane coupling
constant).\cite{quasi2dafm}
However, in the high-temperature paramagnetic phase the
vertical coupling is tiny, i.e., only given by the effective $\Jzz \propto \Jz^4$.
Upon cooling from high $T$ this results in an instability of the
paramagnetic phase at a temperature which is smaller by a factor of 4 compared
to the above estimate.
This argument indicates that the finite-temperature transition could be
discontinuous, but precursor effects of the order-from-disorder
mechanism may counteract, rendering the transition continuous.
A reliable self-consistent treatment of this problem appears difficult and
is beyond the scope of the paper.

In any case, it is clear that the N\'eel temperature will strongly depend on
the frustrated inter-layer coupling $\Jz$.
This applies not only to the situation with O(3) symmetry, but also to the O(2) case:
here, the ordering temperature without vertical coupling
does not vanish, but is given by the Kosterlitz-Thouless temperature $\TKT$,
and $\Jz$ leads to a logarithmic enhancement of $\TN$ compared to $\TKT$.
If the inter-layer coupling is the sum of an unfrustrated and
a frustrated part, then both will contribute to $\TN$ -- this is likely
the scenario relevant to \bacu.


\section{Case (B): Inequivalent layers}
\label{sec:ineq}

We now turn to the case where the ideal bct structure is distorted such
that the unit cell is enlarged to contain multiple dimers.
This modification is suggested by the results of zero-field
inelastic neutron scattering on \bacu, where indications for at least two
modes were found.\cite{rueggbacusio}

As discussed in Sec.~\ref{sec:modineq}, we restrict our calculations
to a situation with two types of inequivalent layers, $A$ and $B$,
which are stacked in an alternating fashion.
(More complicated symmetry-breaking patterns only lead to
quantitative modifications.)
We will consider both the cases of perfect and imperfect frustration,
the latter modeled by unequal vertical couplings along the two diagonals,
see Sec.~\ref{sec:modineq}.

The different microscopic couplings $J_{A,B}$ and $J'_{A,B}$
in the even and odd layers then translate into unequal
order parameter masses $m_{\phi A,B}$ and velocities $c_{A,B}$ in Eq.~\eqref{phi4},
whereas imperfect frustration implies finite $\kappa$ in Eq.~\eqref{sab}.
(In the following we will assume $m_{\phi A} < m_{\phi B}$.)

Numerical results will be obtained from the bond-operator theory in the
harmonic approximation, applied to both the disordered and ordered
phases.
For simplicity, we will not take into account any effects beyond quadratic
terms; this implies that we miss, e.g., effects of the collinear coupling between
the two condensates. This will affect some low-energy properties and will be noted
below, but the gross features of the results can be expected to be correct.

\subsection{Paramagnetic phase: Dispersions and reduced dimensionality}
\label{sec:ineqdisp}

In the situation of two positive triplon masses, the thermodynamic and magnetic
properties of the system will be governed by the smaller of the two gaps.
The primary signature of the doubled unit cell is the presence of
two triplet modes at fixed wavevector (instead of one).
Let us focus on the dispersion of these modes, in a regime
of small to moderate vertical couplings $\Jz$ (neglecting the tiny $\Jzz$).
Then, the in-plane dispersion is conventional and dominated by the $J$, $J'$ values.
At fixed ${\vec q}_\parallel$, the energy of the two modes differs by $\Delta\omega_q$.
For $\Jz \ll \Delta\omega_q$, the vertical dispersion arises only in
{\em second} order in $\Jz$ (!), i.e., $\propto \Jz^2 \cos (2q_z)/\Delta\omega_q$.
In contrast, for $\Jz \gg \Delta\omega_q$ (which includes the case of equivalent planes)
the vertical dispersion is $\propto \Jz$ as usual.

\begin{figure}[t]
\epsfxsize=3.4in
\centerline{\epsffile{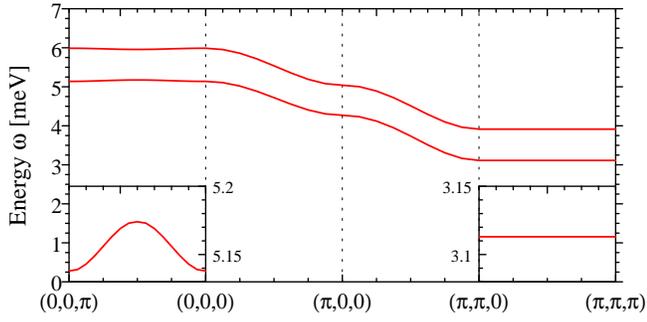}}
\caption{
Zero-field triplon dispersion for inequivalent layers with
perfect frustration, calculated using the harmonic bond-operator approximation.
Parameter values are:
$J_A=4.27$ meV, $J_B=5.04$ meV, $J_A'=J_B'=0.5$ meV, $-\Jtz=\Jz=0.1$ meV.
Compared to the equivalent-layer case, the vertical dispersion at the unfrustrated point,
i.e., along $(0,0,q_z)$, is both quenched and modified in shape,
from $\propto \Jz \cos q_z$ to $\propto \pm \Jz^2 \cos 2 q_z$.
}
\label{fig:disp12_h0}
\end{figure}

Hence, inequivalent layers present an alternative mechanism for {\em dimensional reduction},
which moreover is effective for all in-plane wavevectors and in the absence of
perfect frustration.
Explicitly: for imperfect frustration, the relevant 3d energy scale [being
the bandwidth along $(\pi,\pi,q_z)$] is now $\propto \Jz^2$ for small $\Jz$.
Note that perfect frustration results in a prefactor of $\cos^2 (q_x/2) \cos^2 (q_y/2)$
in front of the bare vertical hopping, which leaves us with the interaction-induced
contribution to the vertical coupling at $q_x=\pi$ or $q_y=\pi$,
which is still of order $\Jz^4$.

\begin{figure}[b]
\epsfxsize=3.4in
\centerline{\epsffile{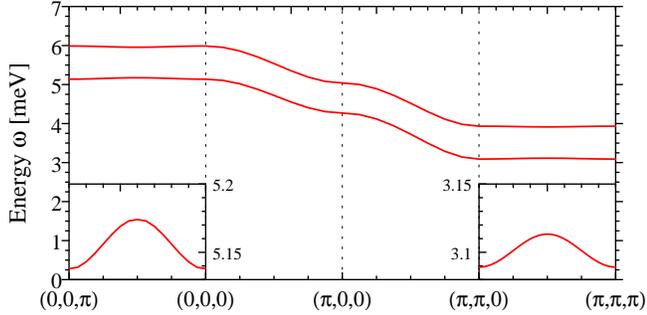}}
\caption{
As in Fig.~\ref{fig:disp12_h0}, but now for inequivalent layers with
{\em imperfect} frustration.
Parameter values are:
$J_A=4.27$ meV, $J_B=5.04$ meV, $J_A'=J_B'=0.5$ meV,
$J_{z1}=0.05$ meV, $J_{z2}=0.15$ meV -- the latter two values are
the couplings along the two inequivalent $\Jz$ diagonals, see
Fig.~\ref{fig:struct}b.
}
\label{fig:disp13_h0}
\end{figure}

Let us make these statements quantitative, by fitting the \bacu\ neutron scattering data
of Ruegg {\em et al.}\cite{rueggbacusio}.
As we model only two inequivalent planes, we take the neutron peaks with smallest and
largest energies and find the (effective) couplings
$J_A=4.27$ meV, $J_B=5.04$ meV, $J_A'=J_B'=0.5$ meV,
being essentially identical to the ones estimated in Ref.~\onlinecite{rueggbacusio}.
Not much is known about the values of the vertical couplings in \bacu.
The simplest assumption is perfect frustration -- a sample result
for the zero-field triplon dispersion for $\Jz=0.1$ meV is shown
in Fig.~\ref{fig:disp12_h0}.
The vertical bandwidth along $(0,0,q_z)$ is roughly 0.04 meV:
compared to the equivalent-layer case, it is suppressed here by a factor of 5,
and it follows $\cos 2 q_z$ instead of $\cos q_z$ (!).
The bandwidth along $(\pi,\pi,q_z)$ vanishes due to the perfect frustration.
(The tiny $\Jz^4$ contribution is not captured in the harmonic approximation.)
For comparison, in Fig.~\ref{fig:disp13_h0} we also show dispersions in a
situation with imperfect frustration,
with couplings along the two inequivalent diagonals of strength
$J_{z1}=0.15$ meV, $J_{z2}=0.05$ meV.
Here, the bandwidth at the unfrustrated point is as before,
but along $(\pi,\pi,q_z)$ we now observe a dispersion of width 0.025 meV.

\begin{figure}[t]
\epsfxsize=3.3in
\centerline{\epsffile{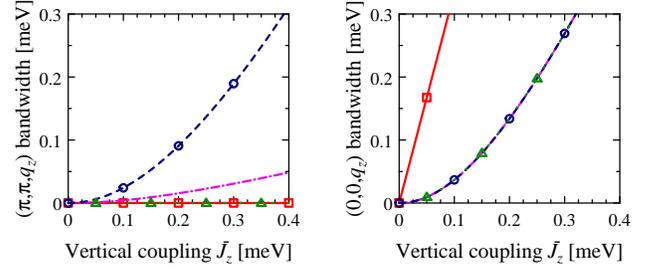}}
\caption{
Bandwidths of the vertical dispersion along $(\pi,\pi,q_z)$ and $(0,0,q_z)$
as function of the average $\bar\Jz$,
for equivalent layers ($J=4.66$ meV) and perfect frustration (solid, squares),
inequivalent layers ($J_A=4.27$ meV, $J_B=5.04$ meV) with perfect frustration (dotted, triangles),
imperfect frustration with $J_{z1}/J_{z2}=3$ (dashed, circles),
and imperfect frustration plus vertical modulation with
$J_{zA1}/J_{zA2}=11/9$, $J_{zB1}/J_{zB2}=19$, ${\bar J}_{zA} = {\bar J}_{zB}$
(dash-dot, no symbols, see also Fig.~\ref{fig:disp14_h0}) below.
In all cases, $J'=0.5$ meV.
}
\label{fig:bw}
\end{figure}

\begin{figure}[b]
\epsfxsize=3.3in
\centerline{\epsffile{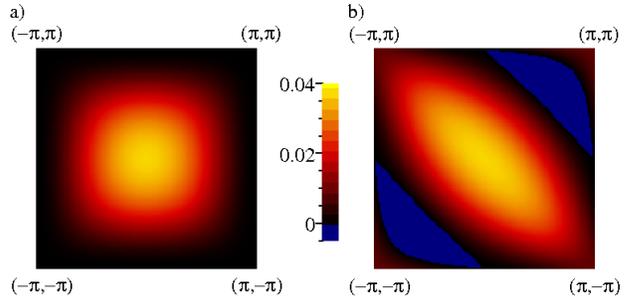}}
\caption{
Vertical dispersion of the lowest mode, $\omega(q_x,q_y,\pi/2)-\omega(q_x,q_y,0)$,
as a function of in-plane momentum ${\vec q}_\parallel$
for inequivalent-layer situations.
a) Perfect frustration, parameter values as in Fig.~\ref{fig:disp12_h0}.
b) Imperfect frustration plus vertical modulation, parameter values as in Fig.~\ref{fig:disp14_h0})
below.
}
\label{fig:bwk}
\end{figure}

In Figs.~\ref{fig:bw},\ref{fig:bwk} we show vertical bandwidths as function of $\Jz$
for the above values of in-plane coupling.
For $\Jz\ll\Delta\omega_q$, these bandwidths scale as
$(J_{zA1}-J_{zA2})(J_{zB1}-J_{zB2})/\Delta\omega_q$ at the frustrated point and
$(J_{zA1}+J_{zA2})(J_{zB1}+J_{zB2})/\Delta\omega_q$ at the unfrustrated point
[with the conventions of Eq.~\eqref{jzdef}].
In Fig.~\ref{fig:bwk} the vertical bandwidths are shown as function of the two-dimensional
in-plane momentum, for the parameter values of Figs.~\ref{fig:disp12_h0}
and \ref{fig:disp14_h0} (below, this case features inequivalent vertical diagonals).

\subsection{Antiferromagnetic phases: Frustrated proximity effect}

The antiferromagnetic phases in the inequivalent-layer case,
both at zero and finite field,
can be nicely discussed using the order-parameter field theory.
Due to the unequal masses for the $A$ and $B$ subsystems,
a situation with $m_{\phi A}<0$, $m_{\phi B}>0$ can occur.
Then, order in the $A$ subsystem is established.

Usually, one would expect that $A$ order induces order in the $B$ subsystem
as well, due to a proximity effect. Technically, a proximity effect arises
from a linear coupling between $\vec\phi_A$ and $\vec\phi_B$
[the $\kappa$ term in ${\cal S}_{\phi AB}$ \eqref{sab}].
However, as detailed in Sec.~\ref{sec:sym}, this term
is forbidden in the presence of perfect frustration, and
hence the proximity effect is absent due to frustration!
The only effect of the $A$ condensate on the $\vec\phi_B$ fluctuations
is via the quartic terms which cause an anisotropy, see
below.

The absence of a proximity effect also implies that the (strong)
order-from-disorder mechanism discussed in Sec.~\ref{sec:eqofd}
is only present in phases with order in both $A$ and $B$ subsystems.

\subsection{Phase diagrams}

\begin{figure}[t]
\epsfxsize=3in
\centerline{\epsffile{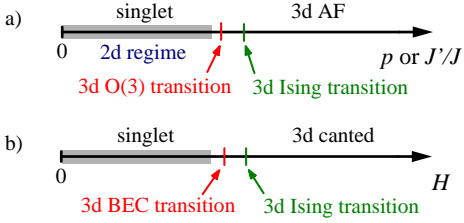}}
\caption{
Schematic $T=0$ phase diagrams for the coupled-dimer model with
inequivalent layers and perfect frustration.
Due to the absence of a proximity effect, there is no coupling between
the condensates on the subsystems $A$ and $B$ (on even and odd layers),
leading to two magnetic transitions.
a) Pressure tuning. b) Field tuning.
}
\label{fig:schemineq}
\end{figure}

\begin{figure}[b]
\epsfxsize=3.5in
\centerline{\epsffile{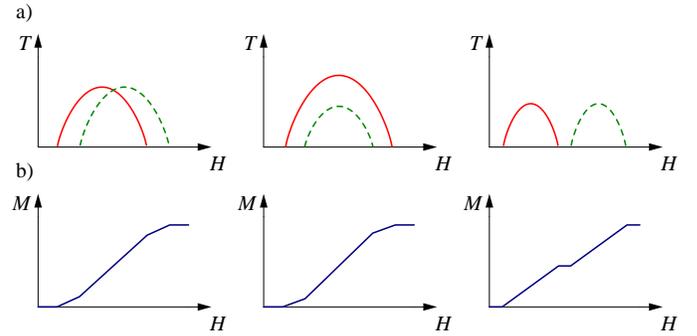}}
\caption{
a)
Schematic overall temperature--field phase diagrams for the coupled-dimer model with
inequivalent layers and perfect frustration, for different
relations between $J_{A,B}$, $J'_{A,B}$.
Solid (dashed) lines show the phase transitions on the $A$ ($B$) subsystem.
b)
Schematic zero-temperature magnetization curves for the three scenarios in a).
For perfect frustration, the slope will change
discontinuously at the transitions.
}
\label{fig:schem_pdgrs}
\end{figure}

Given the absence of a proximity effect in the perfectly frustrated case,
two distinct phase transitions will occur for the $A$ and $B$ subsystems.
Depending on the relation between the parameters $J_{A,B}$, $J'_{A,B}$,
different phase diagrams may be realized, see
Figs.~\ref{fig:schemineq},\ref{fig:schem_pdgrs}.

There will be primary transitions between a disordered and a partially
symmetry-broken phase, and secondary transitions between partially and fully
symmetry-broken phases. Both types transitions extend to finite temperatures
and will be visible as singularities in thermodynamic quantities.
Breaking the frustration will turn the secondary transitions into crossovers,
see Sec.~\ref{sec:ineqhpd} for explicit numerical results.

We believe that the presence or absence of these secondary transitions
allows a clear-cut experimental distinction between perfect and imperfect
frustration in the situation (B) of inequivalent layers.

\subsection{Zero-field phase transitions}

For perfect frustration,
upon increasing the ratios $J'/J$ (e.g. by applying pressure)
either $\vec\phi_A$ or $\Psi$ may condense first.
If the difference between the masses, $m_{\phi B}- m_{\phi A}$,
is larger than binding energy of two $A$ and $B$
triplons (which is set by $\Jfz$), then the first transition
is the magnetic ordering of $\vec\phi_A$.
This transition is in the standard O(3) universality class and has $z=1$.
The dimensional crossover between 3d and 2d critical behavior is set by
$\eta'_A \sim \Jz^4$, as discussed in Sec.~\ref{sec:eqppd}.

Ordering of $A$ introduces a mass anisotropy into the action for $\vec\phi_B$
through the term $u_1 (\langle\vec\phi_A\rangle \cdot \vec\phi_B)^2$, which
is of easy-plane (easy-axis) type for $u_1>0$ ($u_1<0$).
Then, further increasing $J'/J$ will lead to ordering of $\vec\phi_B$,
with a quantum phase transition in the O(2) (Ising) universality class with $z=1$.
Recall that $u_1<0$ is the generic situation in our coupled-dimer model, leading
to a secondary Ising transition.
The crossovers near the second transition are somewhat more complicated:
with increasing energy,
we go from 3d Ising to 2d Ising to 2d Heisenberg, with the two crossover
scales set by $\eta_B' \propto \Jz^4$ and $u_1\langle\vec\phi_A\rangle^2 \propto \Jz^2$.

For imperfect frustration there is only one magnetic transition
in the O(3) universality class, and the secondary transition is smeared out.
The energy scale relevant to the dimensional crossover from 3d to 2d is now
determined by the ``bare'' triplon dispersion near the minimum wavevector
$(\pi,\pi,0)$ (i.e. interaction effects are subleading).
As shown in Sec.~\ref{sec:ineqdisp}, this scale is proportional to $\Jz^2$
(i.e. much larger than in the perfectly frustrated case).

\subsection{Field-driven phase transitions}
\label{sec:ineqhpd}

\begin{figure}[t]
\epsfxsize=3.3in
\centerline{\epsffile{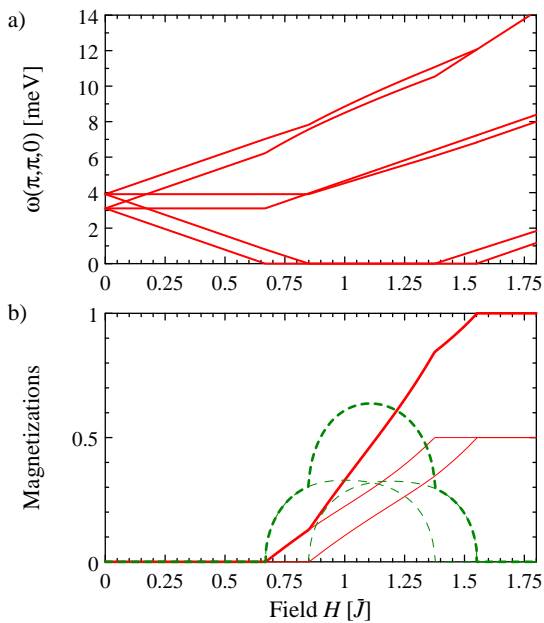}}
\caption{
a) Triplon energy gaps and
b) uniform (solid) and staggered (dashed) magnetization
as in Fig.~\ref{fig:mag11},
but for inequivalent layers with perfect frustration --
the splitting of the phase transitions is clearly visible.
Parameter values are:
$J_A=4.27$ meV, $J_B=5.04$ meV, $J_A'=J_B'=0.5$ meV, $-\Jtz=\Jz=0.1$ meV.
The field is displayed in units of the average $\bar{J} = 4.66$ meV.
The thin lines in b) show the individual contributions
of the A and B subsystems (i.e. even and odd layers) to the
magnetizations.
The presence of two Goldstone modes in the intermediate field range is
an artifact of the harmonic approximation: inclusion of interaction effects
would split the two modes.
}
\label{fig:mag12}
\end{figure}

Similar to the zero-field transition, the field-driven transitions
at $H_{c1}$ and $H_{c2}$ are split in the presence of perfect frustration
due to the absence of a proximity effect.
The primary $A$ ordering transition is in the dilute Bose gas (i.e. BEC) universality
class with $z=2$.
At $H_{c1}$, the dimensional crossover between 3d and 2d critical behavior is
again set by the scale $\eta'_A$.
(As discussed in Sec.~\ref{sec:eqhpd}, this scale vanishes at $H_{c2}$,
if the Hamiltonian has no bare $\Jzz$.)

Once the A subsystem is ordered,
the $u_1$ term introduces an easy-axis anisotropy for the remaining $\vec\phi_B$
U(1) degrees of freedom (irrespective of the sign of $u_1$).
Hence, the secondary transition at $H_{c1}'$ is in the Ising universality class with $z=1$.
(The change from $z=2$ to $z=1$ can be understood, e.g., in terms of triplon modes:
the easy-axis term couples $\tau_+$ and $\tau_-$, and the resulting $2\times 2$ matrix
diagonalization results in a linearly dispersing mode at criticality.
The easy-axis term is neglected in the harmonic bond-operator calculation.)
The crossover with increasing energy is now from 3d Ising to 2d Ising
to 2d BEC, with crossover scales as above.

\begin{figure}[t]
\epsfxsize=3.3in
\centerline{\epsffile{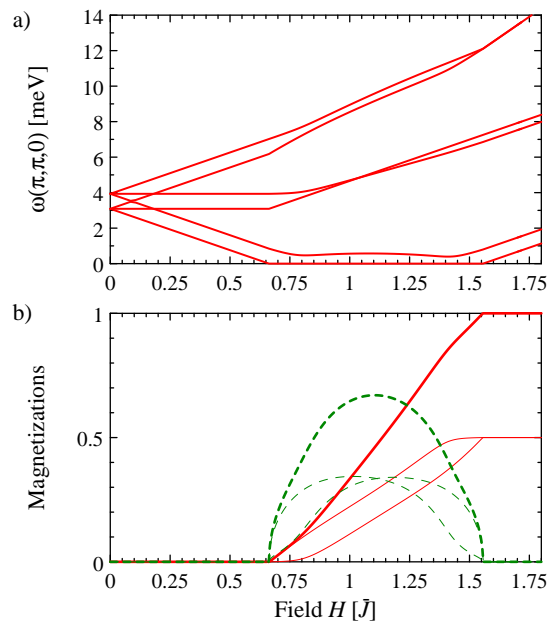}}
\caption{
a) Triplon energy gaps and
b) uniform (solid) and staggered (dashed) magnetization
as in Fig.~\ref{fig:mag12},
but now for inequivalent layers with {\em imperfect} frustration --
here the secondary phase transitions are smeared out due to the
proximity effect.
Parameter values are:
$J_A=4.27$ meV, $J_B=5.04$ meV, $J_A'=J_B'=0.5$ meV,
$J_{z1}=0.05$ meV, $J_{z2}=0.15$ meV -- the latter two values are
the couplings along the two inequivalent $\Jz$ diagonals, see
Fig.~\ref{fig:struct}b.
}
\label{fig:mag13}
\end{figure}

This discussion implies that, for perfect frustration, there will be a field range
$H_{c1}< H < H_{c1}'$ where the $B$ layers are still disordered.
In this regime, the $B$ layer magnetization is zero (exponentially small) at $T=0$ (low $T$),
respectively.
This will only change at the secondary transition which occurs where
the $B$ gap as function of the field reaches zero, leading to a kink
e.g. in the total magnetization, see Figs.~\ref{fig:schem_pdgrs} and \ref{fig:mag12}.
Note that this kink will also be present at finite $T$,
due to the power-law onset of the secondary order parameter at the transition.
The distance between the two transitions is essentially given by the
difference in the zero-field gaps of the two modes $\vec\phi_A$, $\vec\phi_B$;
small corrections arise from the density interactions between the two condensates.

As above, for imperfect frustration, the proximity effect will smear out the
secondary transition, and with it the magnetization kink.
The energy scale for the dimensional crossover is determined
by the bare dispersion of the lowest triplon and scales as $\Jz^2$.

Magnetization curves obtained using the bond-operator method are shown
in Figs.~\ref{fig:mag12} and \ref{fig:mag13}, for perfect and imperfect frustration,
respectively.
For imperfect frustration, the magnetization $m_B$ of the $B$ subsystem is now
finite above $H_{c1}$, but depends in a non-linear fashion on the
$A$ magnetization $m_A$ (or the field).

\subsection{Larger unit cells}

For unit cells containing more than two dimers, the number of triplon modes
increases accordingly, but many qualitative features of our analysis remain
valid.
(i) There will be at maximum {\em two} distinct condensates, not coupled
by a proximity effect in the case of perfect frustration.
Hence, the number of separate sharp ordering transitions will not exceed two,
although the number of mode gaps is larger, the reason being a proximity
effect due to unfrustrated couplings.
(ii) The bandwidth of the vertical dispersion for $N$ inequivalent layers
will in principle be determined by $N$-th order perturbation theory;
depending on the relation between the $\Jz$ and the mode energy
differences $\Delta\omega_q$, bandwidths between $\propto\Jz$ and $\propto\Jz^N$ are
possible.


\section{Application to \bacu}
\label{sec:bacu}

After having collected information about the different scenarios,
we finally come to a more detailed discussion of the behavior of \bacu.
First we try to summarize important experimental results:
(i) The phase boundary near $H_{c1} = 23.2$ T follows the 2d power law with shift
exponent $\psi=1$ down to 30 mK.\cite{sebastian,nmr0}
(ii) Zero-field neutron scattering shows multiple modes, implying
inequivalent dimers.\cite{rueggbacusio}
Together with the structural distortion below 100\,K this suggests
inequivalent layers.
Further, the vertical dispersion of the modes seems tiny ($<0.1$ meV)
even at wavevectors away from the frustrated point.
(iii) NMR experiments show the presence of two inequivalent Cu sites
above $H_{c1}$, with significantly different, but non-zero magnetizations.\cite{nmr}
(iv) The magnetization curve is almost linear above $H_{c1}$.\cite{bacusio}
(v) Small spin anisotropies arise from magnetic dipolar interactions,
but those are of order 10\,mK only.\cite{bacuaniso}

Points (ii) and (iii) rule out a perfect bct lattice structure with a
single-dimer unit cell, Sec.~\ref{sec:eq}.
In the following we therefore assume the inequivalence of even and odd layers,
i.e., our scenario (B) of Sec.~\ref{sec:ineq}.
Further, we will ignore the possible tendency towards
incommensuration, i.e., we stick to  layers with $(\pi,\pi)$ ordering.
We are aware that this may not fully account for all details relevant to
the material.
(Note that a full characterization of the low-temperature lattice structure is still
open.)

Within the scenario of inequivalent layers, perfect frustration implies
two distinct condensates without proximity effect. This would invariably
lead to two transitions as function of field (even at finite $T$),
with a kink in the magnetization curve.
The second transition field can be estimated from the zero-field gaps\cite{rueggbacusio}
to be around $H_{c1}' \approx 27$ T.
In the field range $H_{c1} < H < H_{c1}'$ the magnetization in the
second (non-ordered) subsystem should be exponentially small.
In contrast, the NMR data\cite{nmr} indicate a sizeable magnetization here
and hence appear inconsistent with the perfect-frustration scenario.
Further, to our knowledge no secondary thermodynamic transition has been
observed near 27 T.
Together, this appears to rule out perfect frustration.

We are left to consider inequivalent layers with imperfect frustration.
Imperfect frustration induces a finite vertical bandwidth for all in-plane wavevectors,
and it causes proximity effect, thereby smearing out the secondary ordering transition.
The finite vertical bandwidth at ${\vec q}_\parallel=(\pi,\pi)$ is the relevant dimensional
crossover scale below which 3d behavior is observed
at the phase transition.
From (i), this scale is known to be less than 30 mK -- this significantly
constrains the unfrustrated component, $J_{z1}-J_{z2}$,
of the inter-layer coupling (Sec.~\ref{sec:ineqdisp}):
it has to be smaller than 0.035 meV
(assuming this coupling to be vertically unmodulated;
note that the square of this value is proportional to the vertical bandwidth).
Taking this value, we have calculated magnetization curves which only display tiny
deviations from the case of perfect frustration, Fig.~\ref{fig:mag12}, i.e.,
a pronounced magnetization kink at a second ``transition'' field $H_{c1}'$, and essentially
zero magnetization in the $B$ subsystem below this field.
Apparently, the proximity effect is too small to be in agreement with
both the NMR and the magnetization data.

\begin{figure}[t]
\epsfxsize=3.4in
\centerline{\epsffile{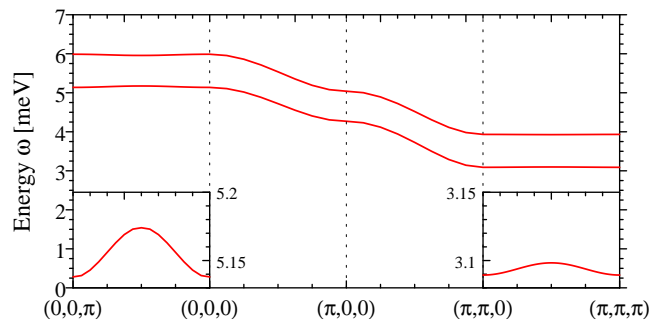}}
\caption{
Zero-field triplon dispersions
for inequivalent layers with partially frustrated and {\em modulated} vertical coupling,
Fig.~\ref{fig:struct}c.
Parameter values are:
$J_A=4.27$ meV, $J_B=5.04$ meV, $J_A'=J_B'=0.5$ meV,
$J_{zA1}=0.09$ meV, $J_{zA2}=0.11$ meV, $J_{zB1}=0.01$ meV, $J_{zB2}=0.19$ meV.
The bandwidth along $(\pi,\pi,q_z)$ is reduced by a factor of 3
as compared to the situation without modulation (Fig.~\ref{fig:disp13_h0}) --
and can be made arbitrarily small by reducing $|J_{zA1}-J_{zA2}|$,
but the magnetization data in the present case are essentially undistinguishable
from the ones in Fig.~\ref{fig:mag13} above.
}
\label{fig:disp14_h0}
\end{figure}

Thus, a 3d scale smaller than 30 mK is not easily compatible with
the large proximity effect. To reconcile these facts, we
propose the inter-layer coupling to be both frustrated and
{\em vertically modulated}, see Fig.~\ref{fig:struct}c.
In such a situation, the unfrustrated components of the inter-layer coupling
can alternate between being weak and strong, rendering a large proximity
effect compatible with a tiny 3d dispersion.
To prove the case, we have calculated dispersions and magnetization data in
a situation with
$J_{zA1}=0.11$ meV, $J_{zA2}=0.09$ meV, $J_{zB1}=0.19$ meV, $J_{zB2}=0.01$ meV,
i.e., the unfrustrated component of the inter-layer alternates strongly,
while the frustrated component is unmodulated.
With this choice, we obtain magnetization data with a sizeable proximity effect
very similar to Fig.~\ref{fig:mag13}, but the vertical dispersion along $(\pi,\pi,q_z)$
[which is given by $(J_{zA1}-J_{zA2})(J_{zB1}-J_{zB2})/\Delta\omega_q$]
is strongly reduced, Fig.~\ref{fig:disp14_h0}.
By reducing $|J_{zA1}-J_{zA2}|$ further, this 3d scale gets arbitrarily small,
without significant changes in the magnetizations.
(Note that the precise shape of the magnetization curve will be influenced
by logarithmic corrections not captured by our harmonic approximation, but the
presence or absence of a rather sharp kink will be unaffected.
As stated above, we never find a field-independent ratio of the two magnetizations,
$m_A/m_B$, and attribute the experimental observation\cite{nmr} of $m_A/m_B\approx 5$
to a coincidence related to proximity effect and logarithmic corrections.)

In summary, within the scenario of modulated vertical couplings
we can easily get consistency between two key observations:
a tiny 3d scale and a large proximity effect.
We note that a modulation of the unfrustrated part of the coupling is required
for both phenomena;
the largest unfrustrated coupling is of order 0.2 meV (Fig.~\ref{fig:disp14_h0}) and
hence relevant for the magnetism not only at lowest $T$.
There are more free parameters in the proposed model:
an additional modulation of the frustrated part will mainly reduce
the bandwidth along $(0,0,q_z)$ -- here future neutron scattering experiments
will help to put bounds on these couplings.
In the absence of further experimental information on the 3d dispersion scales
we refrain from a more detailed fitting of the experimental data.

Let us finally comment on the transition temperature $\TN$.
In a general quasi-2d XY magnet with some (unfrustrated) inter-layer coupling $\Jz$,
$\TN$ will sensitively depend on $\Jz$:
For $\Jz\to 0$, $\TN$ is given by the Kosterlitz-Thouless temperature $\TKT$,
but a finite $\Jz$ will induce a strong (logarithmic) enhancement to that, due to the
exponentially large correlation length above the 2d Kosterlitz-Thouless
transition.
For a partially frustrated inter-layer coupling, both the unfrustrated and frustrated
parts will contribute to $\TN$, but the detailed behavior may be complicated
(see the discussion in Sec.~\ref{sec:tn}),
and a quantitative calculation of $\TN$ is beyond the scope of the paper.
Applied to \bacu, we believe that the rather small value of the
unfrustrated 3d dispersion scale ($\lesssim 30$ mK) may well be compatible with the observed $\TN$
(which is roughly twice as large as the expected $\TKT$), as the frustrated
part of the inter-layer coupling is likely significantly larger ($\lesssim 0.5$ K)
and enhances $\TN$ as well.


\section{Conclusions}

We have studied quantum dimer magnets on layered lattices of the bct type.
Guided by experimental results on \bacu, we have discussed two
distinct routes towards magnetism with reduced dimensionality:
(A) an ideal bct lattice with perfect frustration, and
(B) a distorted lattice with inequivalent layers and possibly broken frustration.
While the situations with perfect frustration are theoretically
fascinating, the material \bacu\ likely falls into class (B) without
perfect frustration.
There, dimensional reduction primarily arises from a layer ``mismatch'':
vertical magnon hopping proceeds via a second-order process of strength
$\Jz^2/\Delta\omega_q$ (where $\Delta\omega_q$ is the magnon energy
difference between even and odd layers).
In addition, we propose that a further reduction of the 3d energy
scale originates from a modulation of $\Jz$ along the $c$ axis, with
the unfrustrated part being small on every second link.
This scenario allows to reconcile the presence of inequivalent
layers\cite{rueggbacusio,nmr}
and the tiny 3d energy scale\cite{sebastian}
with the rather large magnetic proximity effect.\cite{nmr}

For the case of perfect frustration, a number of theoretical challenges
are left for future work:
Those include (i) a more detailed study of the dimensional
crossover behavior in the quantum critical regime,
and (ii) the fate of the order-from-disorder mechanism
at elevated temperatures and the nature of the finite-temperature transition,
including a reliable estimate of the N\'eel temperature.


\acknowledgments

We thank C. Batista, I. Fischer, M. Garst, V. N. Kotov,
F. Mila, J. Mydosh, R. Stern
and in particular
B. Normand, A. Rosch, Ch. R\"uegg, and J. Schmalian
for discussions,
furthermore J. Sirker and O. Sushkov for pointing out the role of
vertex corrections in the bond-operator approach in finite fields,
Ch. R\"uegg for sharing unpublished experimental data,
and C. Vojta for assistance and support.
This research was supported by the DFG through
Research Unit 960 and SFB 608.


\appendix
\section{Bond operators beyond the harmonic approximation}
\label{kotovapp}

In this appendix we present details of the paramagnetic bond-operator approach
introduced in Sec.~\ref{sec:kotov} -- this is a generalization of the formalism of
Ref.~\onlinecite{kotov} to finite fields. An additional generalization to finite
temperature is possible as well, but we shall not follow that here.

The Hamiltonian is given by $\mathcal{H}=\mathcal{H}_2+\mathcal{H}_4+\mathcal{H}_U$,
with $\mathcal{H}_2$ and $\mathcal{H}_U$ given in (\ref{H2}) and (\ref{HU}),
respectively. The quartic term $\mathcal{H}_4 = \mathcal{H}_{4\parallel} + \mathcal{H}_{4z}$ is
\begin{eqnarray}
\mathcal{H}_{4\parallel} &=& \frac{J'}{2} \sum_{\langle ij\rangle n}
h_4(in;jn),\nonumber\\
\mathcal{H}_{4z} &=&         \frac{J_{4z}}{4}\sum_{i\Delta n} h_4(in;i+\Delta,n+1),
\label{h4z}
\end{eqnarray}
with dimer site indices $(in)$ as in Eq.~(\ref{H}),
$J_{4z} = J_z^{11} + J_z^{22} + J_z^{12} + J_z^{21}$,
and $h_4$ given by
\begin{eqnarray}
h_4(i;j)
&=&
\Big[
t_{j0}^\dagger t_{i0}^{\phantom\dagger}
\left(t_{i+}^\dagger t_{j+}^{\phantom\dagger}+t_{i-}^\dagger
t_{j-}^{\phantom\dagger}\right) \\
&&\!\!-
t_{i0}^{\phantom\dagger} t_{j0}^{\phantom\dagger}
\left(t_{i+}^\dagger t_{j-}^\dagger+t_{i-}^\dagger t_{j+}^\dagger\right)
+\mathrm{h.c.}
\Big]
\nonumber \\
&&\!\!+
\left(t_{i+}^\dagger t_{i+}^{\phantom\dagger}-t_{i-}^\dagger t_{i-}^{\phantom\dagger}\right)
\left(t_{j+}^\dagger t_{j+}^{\phantom\dagger}-t_{j-}^\dagger t_{j-}^{\phantom\dagger}\right).
\nonumber
\end{eqnarray}
In principle, there is also a cubic term,
\begin{equation}
\mathcal{H}_{3z}\!\!=\!\!\sum_{i\Delta n}\!\left[\frac{J_{3z}^+}{4}h_3(in;i\!+\!\Delta,n\!+\!1)
+\frac{J_{3z}^-}{4}h_3(i\!+\!\Delta,n\!+\!1;in)\right]\!,
\label{h3z}
\end{equation}
with $J_{3z}^\pm\!=\!J_z^{11}\!-\!J_z^{22}\!\pm\!(J_z^{12}\!-\!J_z^{21})$ and
\begin{eqnarray}
h_3(i;j)&=&
\sum_{i\Delta n}\left(
(t_{i0}^{\phantom\dagger}+t_{i0}^\dagger)(t_{j-}^\dagger t_{j+}^{\phantom\dagger}
-t_{j+}^\dagger t_{j-}^{\phantom\dagger})\right.\\
&&\left.+\left[(t_{i+}^{\phantom\dagger}+t_{i-}^\dagger)(t_{j0}^\dagger t_{j-}^{\phantom\dagger}
-t_{j+}^\dagger t_{j0}^{\phantom\dagger})+\mathrm{h.c.}\right]
\right).\nonumber
\end{eqnarray}
We neglect it as its inclusion would not qualitatively affect our conclusions:
Because of the geometric frustration, self-energy contributions deriving from
$\mathcal{H}_{3z}$ which are $\propto J_{3z}^2$ (Fig.~\ref{fig:dgr2}d)
give no $q_z$-dependent correction to the triplon dispersion for
$\vec q_\parallel=(\pi,\pi)$.

The effect of the local and non-retarded interaction ${\cal H}_U$ can be captured
by a ladder summation, leading to a four-point vertex
\begin{equation}
\label{Gamma}
\Gamma_{\alpha\beta,\alpha\beta}({\vec k},\omega)=
-\left(\frac{1}{N}\sum_{\vec p}\frac{
u_{{\vec p}}^2 u_{{\vec k-\vec p}}^2}
{\omega-\omega_{\vec p\alpha}-\omega_{\vec k- \vec p,\beta}}\right)^{-1}
\end{equation}
where $N$ is the number of lattice sites.
Here, all anomalous scattering vertices have been neglected, which is justified
if the number of triplet bosons is small (Brueckner approximation).
(Note that the Bogoliubov coefficients $u$,$v$ do not depend
on the Zeeman index, see below.)

To leading order, the normal self energy from ${\cal H}_U$ is given by the
sum of Hartree and Fock diagrams (Fig.~\ref{fig:dgr2}a):
\begin{eqnarray}
\label{Sigma_U}
\Sigma_\alpha^U({\vec k},\omega)&=&
\Sigma_{\alpha\alpha}({\vec k},\omega) +
\sum_\beta \Sigma_{\alpha\beta} ({\vec k},\omega)\,, \nonumber\\
\Sigma_{\alpha\beta} ({\vec k},\omega) &=&
\frac{1}{N}\sum_{\vec q}
v_{{\vec q}\beta}^2\Gamma_{\alpha\beta,\alpha\beta}({\vec k + \vec q},\omega-\omega_{{\vec q}\beta}).
\end{eqnarray}
For the bct lattice,
this self-energy will induce a triplon dispersion along $(\pi,\pi,q_z)$ of order $\Jz^6$,
with the leading process being the one in Fig.~\ref{fig:dgr1}b2
(the scattering amplitude from ${\cal H}_U$ is of order unity).

In addition to the hard-core interaction ${\cal H}_U$, we take into account
the quartic $t$ terms in ${\cal H}_4$ in a mean-field approximation,
Fig.~\ref{fig:dgr2}b.
This gives additional self energies:
\begin{eqnarray}
\!\!\!\!\!\!\!\!\!\Sigma_\alpha^n(\vec k) &\!\!\!=&\!\!\!
\frac{2}{N}\sum_{\vec q}
(J' \gamma_{\vec k\parallel}\gamma_{\vec q\parallel}\!+\!J_{4z} \gamma_{\vec kz}\gamma_{\vec qz})
\sum_{\beta\neq\alpha} v_{{\vec q}}^2, \nonumber\\
\!\!\!\!\!\!\!\!\!\Sigma_\alpha^a(\vec k) &\!\!\!=\!\!\! &
-\frac{2}{N}\sum_{\vec q}
(J' \gamma_{\vec k\parallel}\gamma_{\vec q\parallel}\!+\!J_{4z} \gamma_{\vec kz}\gamma_{\vec qz})
\sum_{\beta\neq\alpha} u_{{\vec q}}v_{{\vec q}}
\label{Sigma_na}
\end{eqnarray}
with $\gamma_{\vec q\parallel}$ and $\gamma_{\vec kz}$ defined in Eq.~(\ref{eq:def_gammas}).
While these mean-field terms do not contribute to the dispersion along $(\pi,\pi,q_z)$,
higher-order processes in ${\cal H}_{4z}$ do so.
Remarkably, the second-order self energy, Fig.~\ref{fig:dgr2}c, turns
out to give the {\em leading} contribution to the vertical dispersion,
scaling with $\Jz^4$ ($\Jz^2$ from the vertices and $\Jz^2$ for vertical
hopping on {\em one} internal line).
Therefore we include a contribution to
$\Sigma_\alpha^n(\vec k)$, which is of second order in ${\cal H}_{4z}$, in addition to the
Hartree-like terms above.
The evaluation of the diagram in Fig.~\ref{fig:dgr2}c gives
(here displayed for $\alpha=+$):
\begin{eqnarray}
\Sigma_+^{n,\mathrm{extra}}(\vec k,\omega)&=&
J^2_{4z}\frac{4}{N^2}\sum_{\vec p\vec q}\gamma^2_{\vec k-\vec q,z}\left[f_{\vec k\vec p\vec q\omega}(+,0,0)\right.\nonumber\\
&&\!\!\!\!\!\!\!\!\!\!\!\!\!\!\!\!\!\!\!\!\!\!\!\!\!\!\!\!\!\!\!\!\!\!\!\!\!
\left.+2f_{\vec k\vec p\vec q\omega}(0,0,-)+2f_{\vec k\vec p\vec q\omega}(+,+,+)+f_{\vec k\vec p\vec q\omega}(+,-,-)\right]\nonumber\\
\end{eqnarray}
with
\begin{equation}
\label{f_func}
f_{\vec k\vec p\vec q\omega}(\alpha,\beta,\gamma)
=
\frac{u^2_{\vec k+\vec p-\vec q}u^2_{\vec q}v^2_{\vec p}}
{\omega-\omega_{\vec k+\vec p-\vec q,\alpha}-\omega_{\vec q,\beta}-\omega_{\vec p,\gamma}}.
\end{equation}

The normal self energies $\Sigma^U$ and $\Sigma^n$ add up,
$\bar\Sigma^n(\vec k) = \Sigma^n(\vec k)+\Sigma^U({\vec k},\omega=0)$.
With the self energies we can define renormalized coefficients
$\tilde A_{{\vec k}}$, $\tilde B_{{\vec k}}$ with
\begin{eqnarray}
\tilde A_{{\vec k}} = A_{\vec k}+ \bar\Sigma^n(\vec k), ~
\tilde B_{{\vec k}} = B_{\vec k}+\Sigma^a(\vec k) .
\end{eqnarray}

Solving the coupled Dyson equations for the normal and anomalous Green's functions
yields the renormalized spectrum for the quasiparticles,
\begin{equation}
\label{sp}
\Omega_{{\vec k}\alpha}=
Z_{{\vec k}} \sqrt{\tilde A_{{\vec k}}^2-\tilde B_{{\vec k}}^2} -\alpha H,
\end{equation}
with a quasiparticle weight
\begin{equation}
\label{zk}
Z_{{\vec k}}^{-1} = 1-\left.\frac{\partial \Sigma^U(\vec k,\omega)}{\partial \omega}
\right|_{\omega=0} ,
\end{equation}
and the renormalized Bogoliubov coefficients:
\begin{eqnarray}
\label{uv}
U^2_{{\vec k}},V^2_{{\vec k}} &=&
\pm\frac{1}{2} + \frac{\tilde A_{{\vec k}}}
{2\sqrt{\tilde A_{{\vec k}}^2-\tilde B_{{\vec k}}^2}} \,, \nonumber\\
U_{{\vec k}} V_{{\vec k}} &=&
- \frac{\tilde B_{{\vec k}}}{2\sqrt{\tilde A_{{\vec k}}^2-\tilde B_{{\vec k}}^2}}.
\end{eqnarray}
A fully self-consistent solution of the problem requires
the substitutions
\begin{equation}
\omega_{{\vec k}\alpha} \rightarrow \Omega_{{\vec k}\alpha}, ~~
u_{{\vec k}} \rightarrow \sqrt{Z_{\vec k}}U_{{\vec k}},~~
v_{{\vec k}} \rightarrow \sqrt{Z_{\vec k}}V_{{\vec k}}
\label{substuv}
\end{equation}
in the equations (\ref{Gamma},\ref{Sigma_U},\ref{Sigma_na},\ref{f_func}).

The reader may wonder about a possible $\alpha$ (spin) dependence of the self energies
and the Bogoliubov coefficients. Indeed, the naive perturbative approach sketched here
results in such a spin dependence, which consequently also yields a renormalization of
the magnetic field (or the $g$ factor).
Physically, such a result is incorrect, as the $z$ component of the total spin is
conserved.
This conservation law can be formally implemented by a corresponding Ward identity,
resulting in appropriate vertex corrections.
Here, a consistent result can be obtained by implementing a finite field through
the replacement $\omega \rightarrow \omega + \alpha H$ everywhere.
This implies that the self-energies need to be evaluated at energy $\alpha H$
instead of zero.
Doing this, the self energies and Bogoliubov coefficients remain $\alpha$
independent.\cite{sirker}
(Practically, only the $\alpha=0$ self energy needs to be evaluated.)

\section{Scattering amplitude for Bethe-Salpeter equation}
\label{bseapp}

We consider a biquadratic coupling between adjacent layers,
i.e. a simplified version of Eq.~\eqref{Hcoll}:
\begin{equation}
{\cal H}_\mathrm{coll}=\frac{J_\mathrm{coll}}{4}\sum_{i\Delta nmm'}
\left(
\vec S_{inm}
\cdot
\vec S_{i+\Delta,n+1,m'}
\right)^2,
\end{equation}
rewrite it in terms of bond operators and extract from the quartic terms the following
contribution to the bare scattering amplitude
\begin{eqnarray}
M^\mathrm{coll}_{\alpha\beta,\gamma\delta}&=&
\frac{J_\mathrm{coll}}{2}\big[
(\delta_{\alpha\gamma}\delta_{\beta\delta}
+\delta_{\alpha\beta}\delta_{\gamma\delta})
\gamma_{\vec p+\vec q,z}\nonumber\\
&&
+(\delta_{\alpha\delta}\delta_{\beta\gamma}
+\delta_{\alpha\beta}\delta_{\gamma\delta})
\gamma_{\vec p-\vec q,z}
\nonumber\\
&&
+(\delta_{\alpha\gamma}\delta_{\beta\delta}
+\delta_{\alpha\delta}\delta_{\beta\gamma})
\gamma_{\vec Qz}
\big]
\end{eqnarray}
where we employ the notation of Ref.~\onlinecite{kotovbound}. This adds to the result
deriving from ${\cal H}_4$ and ${\cal H}_U$,
\begin{eqnarray}
M^\mathrm{other}_{\alpha\beta,\gamma\delta}&=&
2(\delta_{\alpha\gamma}\delta_{\beta\delta}
-\delta_{\alpha\beta}\delta_{\gamma\delta})
(J'\gamma_{\vec p+\vec q,\parallel}+J_{4z}\gamma_{\vec p+\vec q,z})\nonumber\\
&&
+2(\delta_{\alpha\delta}\delta_{\beta\gamma}
-\delta_{\alpha\beta}\delta_{\gamma\delta})
(J'\gamma_{\vec p-\vec q,\parallel}+J_{4z}\gamma_{\vec p-\vec q,z})\nonumber\\
&&
+U(\delta_{\alpha\delta}\delta_{\beta\gamma}
+\delta_{\alpha\delta}\delta_{\beta\gamma}).
\end{eqnarray}
Altogether, the scattering amplitude in the singlet ($S=0$) channel
is
\begin{eqnarray}
M^{(0)}&=&\frac{1}{3}\delta_{\alpha\beta}\delta_{\gamma\delta}
(M^\mathrm{other}_{\alpha\beta,\gamma\delta}+
M^\mathrm{coll}_{\alpha\beta,\gamma\delta})\nonumber\\
&=&-4J'(\gamma_{\vec p+\vec q,\parallel}+\gamma_{\vec p-\vec q,\parallel})
-4J_{4z}(\gamma_{\vec p+\vec q,z}+\gamma_{\vec p-\vec q,z})\nonumber\\
&&+2U+J_\mathrm{coll}(2\gamma_{\vec p+\vec q,z}+2\gamma_{\vec p-\vec q,z}
+\gamma_{\vec Qz}).
\end{eqnarray}
Interestingly, the quartic scattering terms arising from $J_{4z}$ and $\Jcoll$
are of essentially identical form (although the Hamiltonians are different).
Consequently, both ${\cal H}_{4z}$ and $\Hcoll$ cause similar triplon
bound states.


\end{document}